\documentclass{aastex631}
\usepackage{graphicx}
\usepackage{epstopdf}
\usepackage{subfigure} 
\usepackage{color}
\usepackage{float}
\usepackage[figuresright]{rotating}
\usepackage{booktabs} 
\usepackage{footnote}
\usepackage{amsmath}
\usepackage{subfigure} 
\usepackage{appendix}
\usepackage{rotating} 
\usepackage{longtable}
\usepackage{threeparttable}
\usepackage{booktabs}
\usepackage{hyperref}
\usepackage{CJKutf8}
\begin{document}
\nolinenumbers  
\title{Identifying Hierarchically Triple Star Systems with Gaia DR3 and LAMOST}


\author[0009-0004-9758-0722]{Tongyu He}
\affiliation{College of Physics Science and Technology, Hebei University, Baoding 071002, China}

\author[0000-0002-2577-1990]{Jiao Li}
\affiliation{Yunnan Observatories, Chinese Academy of Sciences (CAS), 396 Yangfangwang, Guandu District, Kunming 650216, P.R. China}

\author[0000-0003-3832-8864]{Jiangdan Li}
\affiliation{Yunnan Observatories, Chinese Academy of Sciences (CAS), 396 Yangfangwang, Guandu District, Kunming 650216, P.R. China}

\author[0000-0003-4829-6245]{Jianping Xiong}
\affiliation{Yunnan Observatories, Chinese Academy of Sciences (CAS), 396 Yangfangwang, Guandu District, Kunming 650216, P.R. China}

\author{Xiaobin Zhang}
\affiliation{National Astronomical Observatories, Chinese Academy of Sciences, Beijing 100101, P.R. China}

\author{Mikhail Kovalev}
\affiliation{Yunnan Observatories, Chinese Academy of Sciences (CAS), 396 Yangfangwang, Guandu District, Kunming 650216, P.R. China}

\author{Qiyuan Cheng}
\affiliation{Yunnan Observatories, Chinese Academy of Sciences (CAS), 396 Yangfangwang, Guandu District, Kunming 650216, P.R. China}

\author{Sufen Guo}
\affiliation{School of Physics and Technology, Xinjiang University, Urumqi 830046, P.R. China}
\affiliation{Yunnan Observatories, Chinese Academy of Sciences (CAS), 396 Yangfangwang, Guandu District, Kunming 650216, P.R. China}

\author{Mingkuan Yang}
\affiliation{National Astronomical Observatories, Chinese Academy of Sciences, Beijing 100101, P.R. China}


\author[0000-0001-5284-8001]{Xuefei Chen}
\affiliation{Yunnan Observatories, Chinese Academy of Sciences (CAS), 396 Yangfangwang, Guandu District, Kunming 650216, P.R. China}

\author[0000-0001-9204-7778]{Zhanwen Han}
\affiliation{Yunnan Observatories, Chinese Academy of Sciences (CAS), 396 Yangfangwang, Guandu District, Kunming 650216, P.R. China}
\affiliation{College of Physics Science and Technology, Hebei University, Baoding 071002, China}

\correspondingauthor{Zhanwen Han}
\email{zhanwenhan@ynao.ac.cn}\\

\begin{abstract}
Triple star systems are critical for understanding stellar dynamics and compact objects in astrophysics, yet confirmed hierarchical triples identified via spectroscopy remain limited. In this study, we identified 23 triple systems by cross-matching the Gaia DR3 non-single star catalog with LAMOST DR10 spectroscopic data, 18 of them are new discoveries. For two well-observed triples, we performed radial velocity curve fitting and light curve analysis to determine their orbital parameters, with inner and outer periods of 1.26 days and 656 days for one triple, and 3.42 days and 422 days for the other. And we compared the results with other studies. We also analyzed the radial velocities (RVs) of these 23 tripls, revealing a range of $\Delta V$ from approximately 40 km/s to 210 km/s. Due to spectral resolution and detection limitations, velocity differences below 45 km/s in binaries and below 90 km/s in the inner binaries of triple systems are challenging to detect. Consequently, our detection range for inner orbital periods is restricted to 0.2–20 days, with the highest efficiency for periods under 10 days. These findings underscore the advantage of spectroscopic observations for identifying triple systems with short inner orbital periods.
\end{abstract}

\section{Introduction} 
\label{sec:intro} 

Nearly half of the stars in the Milky Way are part of binary, triple, or higher-order systems. These multi-star systems are essential for researching stellar formation, evolution, and galactic dynamics. By analysing period distributions, mass ratios and other characteristics of triple systems, we can better understand the processes behind star formation and evolutionary processes. Studying these systems also sheds light on the internal dynamics and interactions within multi-star systems, offering key insights into the formation of more complex stellar configurations \citep{2008eggleton,He2023}.

Hierarchical triple star systems typically consist of a close binary pair with a third star orbiting the binary’s center of mass at a large distance. This configuration minimizes the gravitational perturbation from the third star on the inner binary, thus organizing the system in a hierarchical manner \citep{aa}. Such an arrangement allows the system to be approximated as a large orbit (the third star around the center of mass of the inner binary) encompassing a smaller orbit (the inner binary itself), facilitating the application of $Kepler$'s laws \citep{He2023}.

In addition to their unique dynamical structure, hierarchical triple systems offer new insights into several longstanding problems in stellar astrophysics. For example, they may show the origin of blue stragglers \citep{Perets2009}, help to explain the formation of Thorne–Zytkow objects \citep{Eisner2022}, and clarify the mechanisms behind recurrent nova outbursts \citep{Knigge2022}.

Hierarchical triple systems can remain stable under certain conditions, ensuring that gravitational interactions between the stars do not lead to system disruption. Key factors for the stability of triple star systems include the hierarchical configuration, significant orbital separation, low orbital eccentricities, appropriate mass ratios, dynamical interactions such as the Kozai-Lidov mechanism and tidal forces \citep{1995ApJ...455..640E,Toonen2017,2019Tokovinin&Moe,aa,GaiaCollaboration2023}. The primary stability criterion involves the period ratio of the outer orbit to the inner orbit, which should exceed a threshold value (around 5) derived from numerical analysis \citep{1995ApJ...455..640E}. This ratio ensures that the gravitational influence of the outer star does not destabilize the inner binary. The stability conditions are described 
by equations that involve the critical initial period ratio  (\(X_{0\text{min}}\)) and the critical initial ratio between the outer orbit’s periastron distance and the inner orbit’s apastron distance (\(Y_{0\text{min}}\)):

\begin{equation}\label{q1}
\left(X_{0}^{\min }\right)^{2 / 3}=\left(\frac{1}{1+q_{\text {out }}}\right)^{1 / 3} \frac{1+e_{\text {in }}}{1-e_{\text {out }}} Y_{0}^{\min } \\
\end{equation}

\begin{equation}\label{q2}
Y_{0}^{\min } \approx 1+\frac{3.7}{q_{\text {out }}^{-1 / 3}}+\frac{2.2}{1+q_{\text {out }}^{-1 / 3}}+\frac{1.4}{q_{\text {in }}^{-1 / 3}} \frac{q_{\text {out }}^{-1 / 3}-1}{q_{\text {out }}^{-1 / 3}+1} \\
\end{equation}


Where \(e_{\text{in}}\) and \(e_{\text{out}}\) are the eccentricities of the inner and outer orbits, respectively, and \(q_{\text{in}} = \frac{M_2}{M_1}\) and \(q_{\text{out}} = \frac{M_3}{M_1 + M_2}\) are the mass ratios of the inner and outer orbits. Here, \(M_1\) and \(M_2\) are the masses of the inner binary, and \(M_3\) is the mass of the third star.


To better understand triple star systems, it is important to accumulate a sufficient sample of systems with orbital parameters. Detecting these systems requires the use of various observational techniques, each methods offering valuable insights into their characteristics and dynamics. Detecting triple star systems can be achieved through various observational techniques, each providing unique insights into the characteristics and dynamics of these systems. Spectroscopic methods, such as radial velocity (RV) measurements and spectral line analysis, can detect the periodic velocity changes due to mutual orbits, revealing the presence of multiple gravitationally bound stars \citep{2010ragha}. \citet{Tokovinin2006} found that photometric methods, including eclipsing systems and variability studies, can indicate the presence of multiple stars through characteristic dips in brightness or system luminosity variations due to mutual eclipses and stellar activity. As noted by \citet{2023Halbwachs}, astrometric techniques, including proper motion and parallax measurements, can reveal deviations in the expected trajectory caused by the presence of additional stars in the system. In addition, high-precision observations and interferometry can directly resolve multiple stars in a triple system if the distances between them are sufficiently large \citep{Hillenbrand2009, Sahlmann2011, 2013ARA, Duch_ne_2013}. While gravitational wave observations are primarily associated with binary systems, \citet{Fabio2017} and \citet{Silsbee2017} indicate that they can also provide indirect insights into the dynamics and masses of stars in triple systems by studying mergers of compact objects or orbital perturbations caused by additional companions. However, unlike direct observational techniques such as spectroscopy or photometry, gravitational waves have not yet offered direct evidence for the existence of a third body in these systems. 

With advancements of observational techniques, the number of known triple star systems has steadily increased. \cite{2018tokovin} identified 201 compact hierarchical triple (CHT) systems with outer orbital periods shorter than 1000 days, and \cite{2008eggleton} compiled a catalog of multiple star systems. Studies also indicate that about 10\% of low-mass stars are part of triple systems, with this fraction rising to nearly 50\% for B-type stars \citep{2008eggleton,2010eggleton,2014AJTokovinin,Evans2011,Sana2012,2022tooten}.

The $Gaia$ mission, especially its third data release (DR3), has enhanced our capacity to discover and study multiple star systems. $Gaia$ DR3 provides highly accurate astrometric data for over a billion stars, including their positions, parallaxes, and proper motions. These data help identify potential multiple star systems by detecting subtle motion patterns indicative of gravitational interactions \citep{GaiaCollaboration2023}. Among the data, a catalog of 135,760 objects with Keplerian orbital solutions has been compiled from detailed astrometric measurements. These findings, along with their related parameters, are documented in the $Gaia$ DR3 non-single star (NSS) catalog, accessible through the Vizier database \citep{Gaia2023}. Although the NSS catalog predominantly focuses on binary systems, it can also uncover higher-order systems, like triple star systems. The Gaia DR3 NSS catalog not only includes multiple systems detected through astrometry, but also incorporates photometric and spectroscopic data.


In recent years, NASA’s Transiting Exoplanet Survey Satellite ($TESS$) has become instrumental in the identification and examination of triple star systems. $TESS$, initially designed to discover exoplanets, is capable of detecting photometric variations indicative of multiple star systems. The precise light curves generated by $TESS$ are useful in recognizing potential triple star systems, which exhibit periodic brightness fluctuations due to stellar interactions or eclipses, providing a basis for further investigation \citep{Ricker2015,Sullivan2015}.

The Large Sky Area Multi-Object Fibre Spectroscopic Telescope (LAMOST) has conducted extensive spectroscopic observations of the northern sky. Its Medium-Resolution Spectroscopy Data Release 10 (DR10) offers a wealth of spectral information, including RVs and other stellar parameters for millions of stars \citep{2012cui,2015luo,Zhao2012}. By processing these spectral data in bulk, we can efficiently identify triple star systems and obtain their RV information. This process not only accelerates the detection of multiple star systems but also enhances our understanding of their dynamic characteristics.

In this study, we combined the $Gaia$ DR3 NSS star catalog with the complete LAMOST-MRS-DR10 dataset to efficiently search for triple star systems. And we used TESS light curve data to derive the parameters for two of these 23 identified hierarchical triple systems. We also analyzed the radial velocities (RVs) for 23 triple systems, as well as the j03 triple system. Our paper is structured as follows: In Section \ref{sec:2}, we analyze the samples obtained from matching the $Gaia$ DR3 NSS catalog with LAMOST DR10 data, using spectral template matching to search for triple star systems. In Section \ref{sec:3}, we confirm 23 triple star systems, 18 of which are newly identified. In Section \ref{sec:4}, we perform cross-correlation function (CCF) spectral analysis, RV fitting, and light curve analysis for the identified triple systems to determine two triple's parameters. Section \ref{sec:6} examines the discussion of RVs for the 23 triple systems from LAMOST and the j03 triple system. Finally, section \ref{sec:7} provides summary and conclusion.

\section{LAMOST Data and Template Matching Methodology} \label{sec:2}


LAMOST is a 4-meter quasi-meridian reflecting Schmidt telescope situated in Hebei, China. It features a 5° field of view equipped with 4,000 optical fibers, enabling simultaneous observations of up to 4,000 targets within a 20 square degree area. The LAMOST-MRS has two arms, with the red arm covering wavelengths from 6300 to 6800 Å and the blue arm spanning 4950 to 5350 Å. We focused on the blue arm spectra to identify candidate triple star systems and measure their RVs, as it offers a greater number of absorption lines \citep{li-2021, Li-2022}.

Our method involves cross-matching samples from the $Gaia$ NSS catalog with all LAMOST-MRS-DR10 spectra to identify a set of non-single star candidates. We then performed cross-correlation function (CCF) fitting on the spectra of the initially obtained 32,305 non-single star candidates to validate triple-star systems.

The classical definition of the CCF for stellar spectra is given by:

\begin{equation}
\text{CCF}(h) = \int_{-\infty}^{+\infty} f(x) g(x + h) \, dx 
\end{equation}
where \( f \) is the normalized spectrum, \( g \) is the normalized template spectrum, and \( h \) is the lag in km/s \citep{2017mer}. The CCF is normalized to range from -1 to +1, where +1 indicates perfect correlation and -1 indicates complete anti-correlation. The CCF method is used in spectroscopy to detect and analyze multiple spectral lines. For example, \cite{2017mer} first proposed this method and detected 11 triple candidates. This method has also been applied in studies such as \cite{li-2021, Li-2022, He2023}.

Our data is derived from the blue arm spectra of LAMOST-DR10, as the blue end contains more absorption lines, making it suitable for detecting multiple candidate lines and measuring their radial velocities (RV). To address the challenge of detecting SB (spectroscopic binary) and ST(spectroscopic triple) in spectra with low signal-to-noise ratios (S/N), we selected data with an S/N greater than 20 \citep{2017fer, li-2021}.

\subsection{The improvement of template matching method}
\label{sec:2.1}

In most studies, the template spectrum used for CCF analysis is the solar spectrum \citep{li-2021, Li-2022, He2023}. We made improvements to this approach. We selected template spectra of different temperatures to match the observed spectra corresponding to various stellar temperatures.

\begin{figure}[ht!]
\centering
\includegraphics[width=11cm]{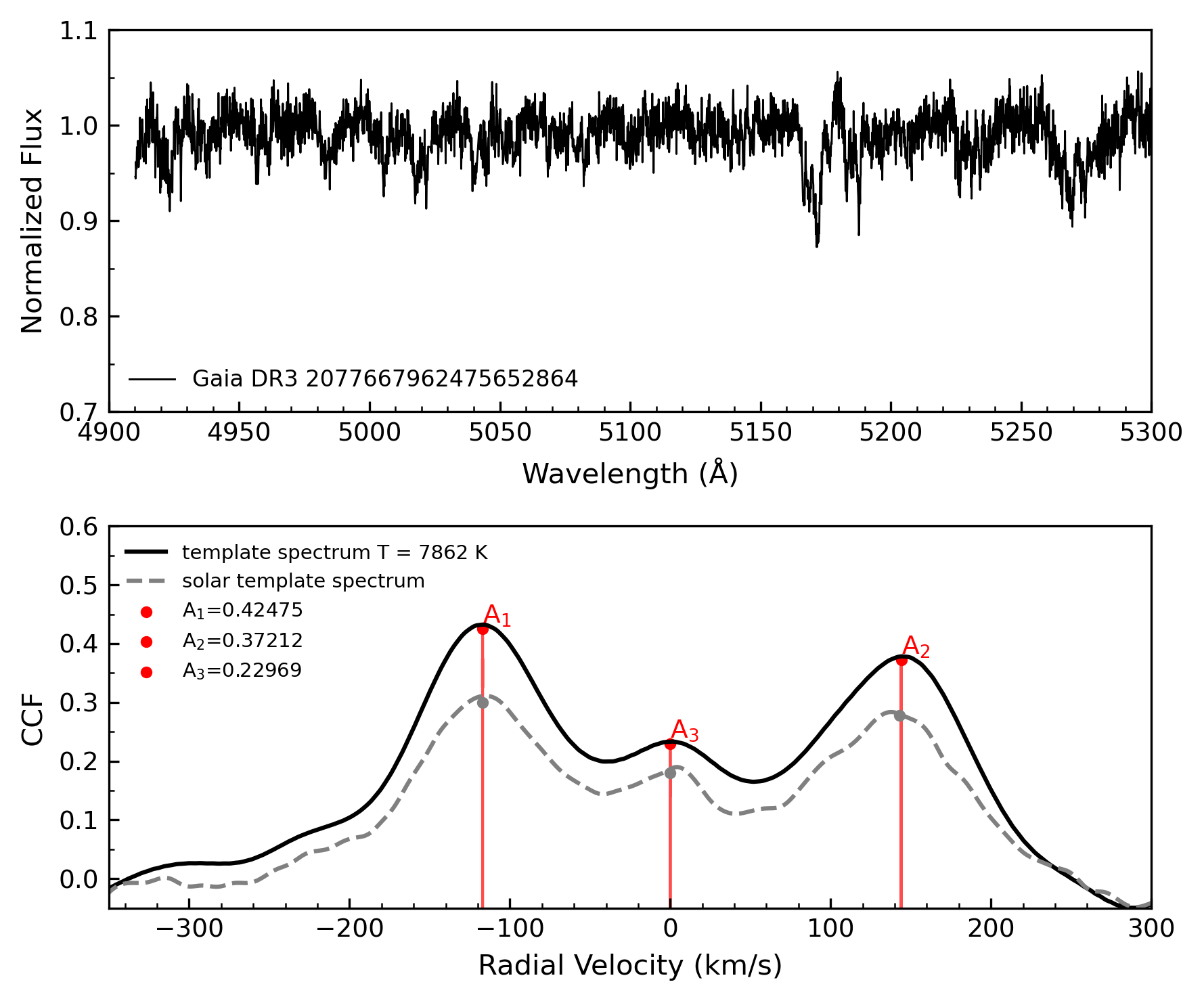}
\caption{Comparison of the CCF peak values for $Gaia$ DR3 2077667962475652864 using different template spectra. In the upper part of the figure shows the normalized spectra. In the lower part of the figure, we present the generated CCF, A1 = 0.53, V1= - 131km/s, A2 = 0.48, V2= 134 km/s, and A3 = 0.41, V3= 19 km/s. The black solid line represents a template spectrum with a temperature of 7862 K, while the grey dashed line represents a CCF generated using a conventional solar spectrum template.
\label{figure 1}}
\end{figure}

As shown in figure {\ref{figure 1}}, using only the solar template spectrum for different types of stellar spectra can result in lower CCF peak values. The figure illustrates the spectrum of a triple-star with a temperature of 7,862K, where the CCF peak obtained with the solar template spectrum is significantly lower than that obtained with a 7,850K template spectrum. We utilized the \texttt{iSpec} software to construct template spectra from LAMOST with wavelengths ranging from 4,900 to 5,300 Å, and temperatures spanning from 4,000K to 8,000K in 50K intervals, while reducing the resolution to 7,500 \citep{2014ispec,2019Blanco}.This generated a series of template spectra. 

Subsequently, we performed CCF fitting using different template spectra based on the stellar temperatures to improve the accuracy 
of the results. As mentioned in \cite{li-2021}, a constraint was set to disregard spectra with CCF values 
less than 0.2. For the SB3 spectra, we retained a purer sample that exhibits clear SB3 characteristics in two or more observations 
\footnote{Other samples will be discussed in next work}.

\section{Identifying 23 Triples From LAMOST}
\label{sec:3}

In section \ref{sec:2}, we cross-matched $Gaia$ NSS samples with LAMOST-MRS-DR10 spectra and performed cross-correlation function (CCF) fitting on non-single star candidates, identifying 23 triple star systems. Five triples had already been identified as SB3 systems ($Gaia$ DR3 508947512847441792, $Gaia$ DR3 3449039200333957248, $Gaia$ DR3 798566091141562240, $Gaia$ DR3 4577886588616414464, $Gaia$ DR3 2077667962475652864) \citep{li-2021}. 

\begin{figure}[ht!]
\centering
\includegraphics[width=11cm]{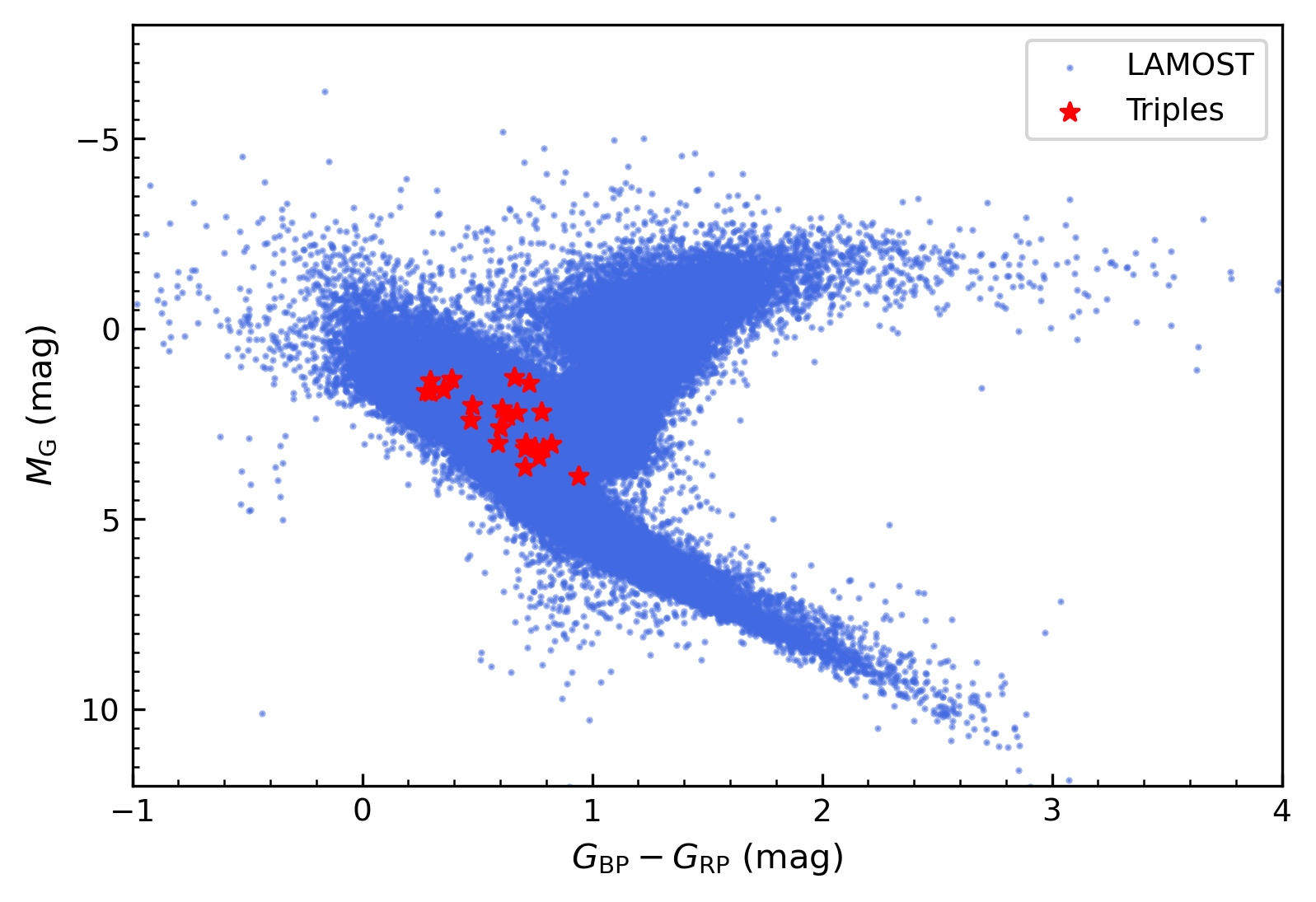}
\caption{The color-magnitude graph of 23 triple-star systems (red stars) determined from $Gaia$ DR3 and LAMOST (blue dots). The x-axis shows the colour difference between $G_\mathrm{BP}$ and $G_\mathrm{RP}$, and the y-axis displays the absolute magnitudes $M_\mathrm{G}$}. All photometric data are from $Gaia$ DR3.
\label{figure 2}
\end{figure}

The color-magnitude diagram for the 23 triples is shown in figure \ref{figure 2}, where the blue dots represent all stars detected with LAMOST medium-resolution spectroscopic (MRS) data, and the red stars indicate the samples we filtered. The plot reveals that the 23 triples are situated close to the main sequence. Through the CCF analysis of the LAMOST-MRS spectra for these samples, we obtained several parameters listed in table \ref{tab1}, which includes the $Gaia$ Source ID, Right Ascension (RA), Declination (DEC), the RV (km/s) of each component in the triple star system, and the amplitude $A$ after CCF processing.


\begin{sidewaystable}
\centering
\tablenum{1}
\caption{The information of triple-star systems.\label{tab1} }
\begin{longtable}{ccccccccccccccccccccc}  
\toprule
Gaia source id & RA (2000) & DEC (2000) & MJD & $\mathrm{RV}_1$ (km/s) & $\mathrm{RV}_{1,\text{err}}$ & $\mathrm{RV}_2$ & $\mathrm{RV}_{2,\text{err}}$ & $\mathrm{RV}_3$ & $\mathrm{RV}_{3,\text{err}}$ & $\mathrm{A}_1$ & $\mathrm{A}_2$ & $\mathrm{A}_3$ \\
\midrule
\endhead
420120476698017792  & 3.02488   & 53.87018  &59183.47292 & -12.0  & 1.21   & 87.0  & 2.5   & 179.0 & 3.11  & 0.73637 & 0.2173  & 0.30318 \\
363358807382928128  & 16.39931  & 35.9877   &58420.60764 & -84.0  & 3.543  & -30.0 & 2.38  & 67.0  & 2.78  & 0.28744 & 0.63114 & 0.28122 \\
307273436508903296  & 18.61908  & 27.88322  &59189.4875 & -77.0  & 3.23   & 20.0  & 3.52  & 94.0  & 2.45  & 0.51685 & 0.24187 & 0.31908 \\
508947512847441792  & 24.50203  & 57.94588  &58088.50833 & -92.0  & 4.19   & -23.0 & 2.74  & 29.0  & 3.26  & 0.45279 & 0.17978 & 0.24136 \\
505328397301344128  & 29.44542  & 57.30957  &58449.61667 & -84.0  & 1.52   & -16.0 & 3.78  & 131.0 & 2.96  & 0.50064 & 0.38804 & 0.32713 \\
442402835747230720  & 54.36964  & 52.19914  &58501.50347 & -14.0  & 3.75   & 58.0  & 3.2   & 158.0 & 3.31  & 0.33321 & 0.30647 & 0.39107 \\
145322211325222272  & 65.91436  & 21.33882  &59243.46875 & -110.0 & 5.37   & -69.0 & 4.97  & -12.0 & 5.52  & 0.20548 & 0.20854 & 0.20376 \\
156838015078007424  & 74.02772  & 30.01244  &58441.70347 & -125.0 & 1.93   & -8.0  & 3.0   & 85.0  & 1.89  & 0.23843 & 0.16384 & 0.37184 \\
213079855906767616  & 77.5241   & 47.51584  &58824.63125 & -55.0  & 2.67   & 4.0   & 1.05  & 79.0  & 1.21  & 0.43457 & 0.3646  & 0.58425 \\
3234943017725508864 & 77.62068  & 2.40336   &58798.79097 & -28.0  & 3.1    & 24.0  & 1.89  & 60.0  & 1.69  & 0.19727 & 0.43785 & 0.37696 \\
3449039200333957248 & 82.22157  & 33.16119  &58534.52014 & -91.0  & 0.66   & 1.0   & 1.59  & 94.0  & 2.12  & 0.31779 & 0.30607 & 0.34055 \\
211453235828167168  & 88.24907  & 49.92759  &59597.65278v & -83.0  & 3.84   & -8.0  & 1.75  & 119.0 & 2.6   & 0.17931 & 0.56398 & 0.19109 \\
689757797684412032  & 134.33757 & 24.11414  &59177.89653 & -46.0  & 2.26   & 48.0  & 1.4   & 155.0 & 0.72  & 0.2308  & 0.24425 & 0.29356 \\
798566091141562240  & 141.26129 & 35.34453  &58088.89653 & -75.0  & 1.805  & 21.0  & 2.87  & 121.0 & 1.76  & 0.41374 & 0.24032 & 0.41734 \\
1498903389706406272 & 209.32449 & 41.8002   &59367.57083 & -85.0  & 1.3    & 0.0   & 4.5   & 70.0  & 1.26  & 0.61006 & 0.24211 & 0.46211 \\
1429211532374616448 & 241.63669 & 54.6728   &58262.64514 & -81.0  & 0.75   & -7.0  & 1.89  & 77.0  & 6.01  & 0.44529 & 0.3918  & 0.29664 \\
4577886588616414464 & 270.26957 & 24.32913  &58626.72014 & -81.0  & 2.96   & -11.0 & 0.93  & 41.0  &0.78  & 0.30679 & 0.46308 & 0.52114 \\
4508809495727468544 & 280.61107 & 13.98528  &59687.8375 & -85.0  & 3.0    & -25.0 & 4.01  & 44.0  & 0.57  & 0.61882 & 0.27672 & 0.45008\\
2067911995796118144 & 308.89664 & 41.77094  &58409.52708 & -112.0 & 0.97   & -12.0 & 3.46  & 66.0  & 0.42  & 0.36734 & 0.18347 & 0.59101 \\
1995477578515779712 & 346.03668 & 52.94115  &58444.50208 & -71.0  & 1.65   & -10.0 & 0.4   & 51.0  & 0.54  & 0.50984 & 0.4423  & 0.54104 \\
1991765489817742208 & 347.79035 & 50.43096  &58444.44861 & -117.0 & 1.48   & -68.0 & 3.74  & 30.0  & 2.33  & 0.55431 & 0.38971 & 0.31471 \\
249662295687401216 & 56.145168 & 50.252774  &59619.46045 &-119.0 & 0.95  & -7.58& 1.73  &139.6& 1.65 & 0.2825& 0.1283&0.3290 \\
2077667962475652864 &292.718018 & 41.922459 &58625.27818 & -71.5 & 1.21 & 6.58& 0.75 & 88.13 & 1.29& 0.5693&0.2954&0.4880 \\
\bottomrule
\end{longtable}
\end{sidewaystable}


\section{Orbital Parameters Of Two Triple Systems} \label{sec:4}
In the previous section, we identified 23 triple star systems, as summarized in table \ref{tab1}. In this section, we extracted spectra for triples with more than six observations. This screening resulted in two triple star systems ($Gaia$ DR3 249662295687401216 and $Gaia$ DR3 2077667962475652864). We discussed the two triples separately.

\subsection{Determination of Orbital Parameters for Gaia DR3 249662295687401216}
\label{sec:4.1}
\subsubsection{Fitting of radial velocity (RV) curve}
\label{sec:4.1.1}

We used the \texttt{laspec} software package \citep{2020zhangbo, 2021zhangbo} to perform CCF analysis on the observed spectra, using the same template spectra as described previously. This method allows us to determine the RVs of each component in the triple-star systems at different BJD times, as shown in figure \ref{figure 3}. See tables \ref{tab4} in the appendix for more details.

\begin{figure}[ht!]
\centering
\includegraphics[width=11cm]{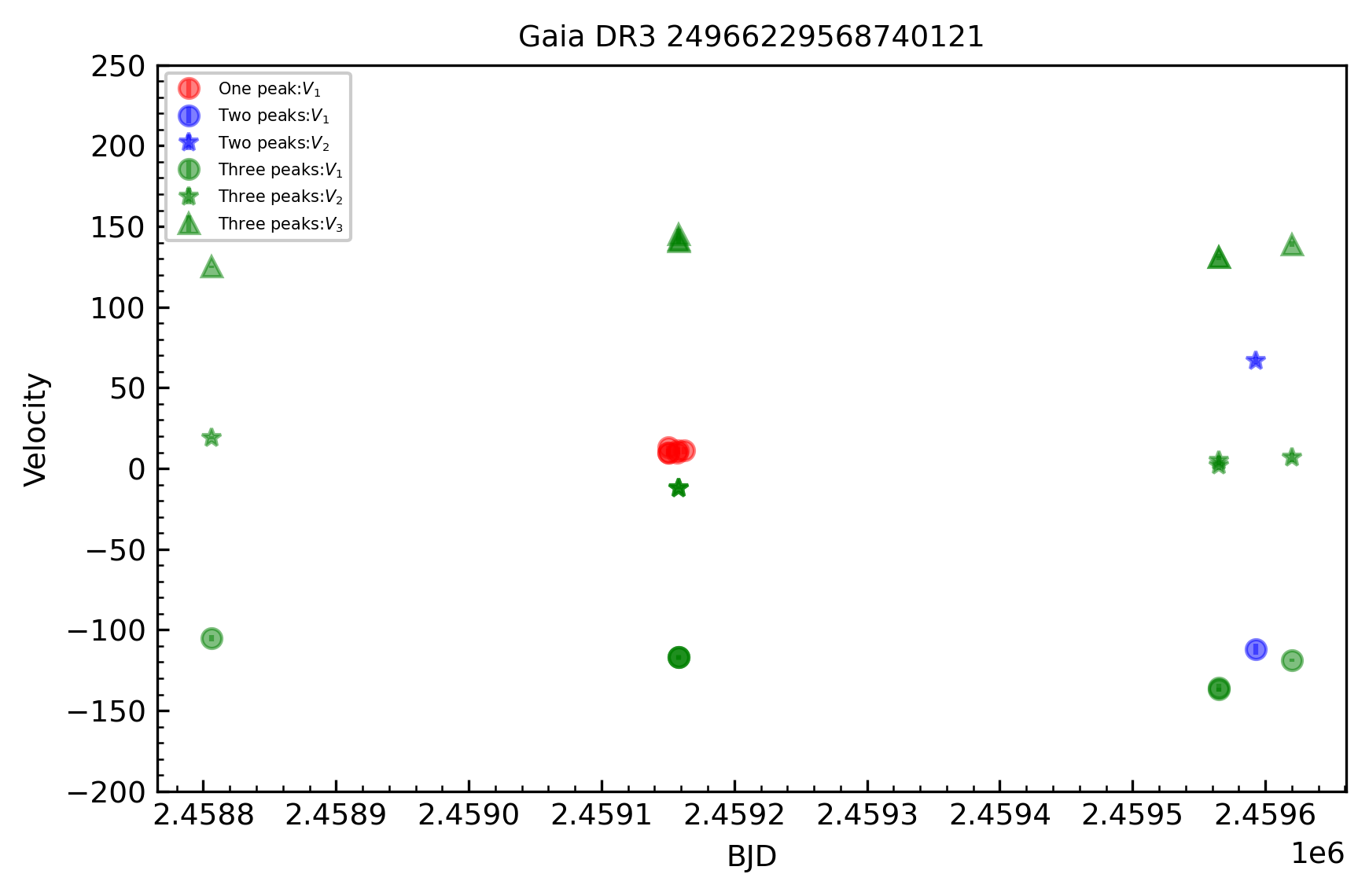}
\caption{RV variations of the triple star system $Gaia$ DR3 249662295687401216 over time. Red circles indicate spectra where CCF analysis detected only one RV. Blue circles and stars represent spectra with two detected RVs, and green circles, stars, and triangles denote spectra with three different RVs.
\label{figure 3}}
\end{figure} 

\begin{figure}[ht!]
\centering
\includegraphics[width=11cm]{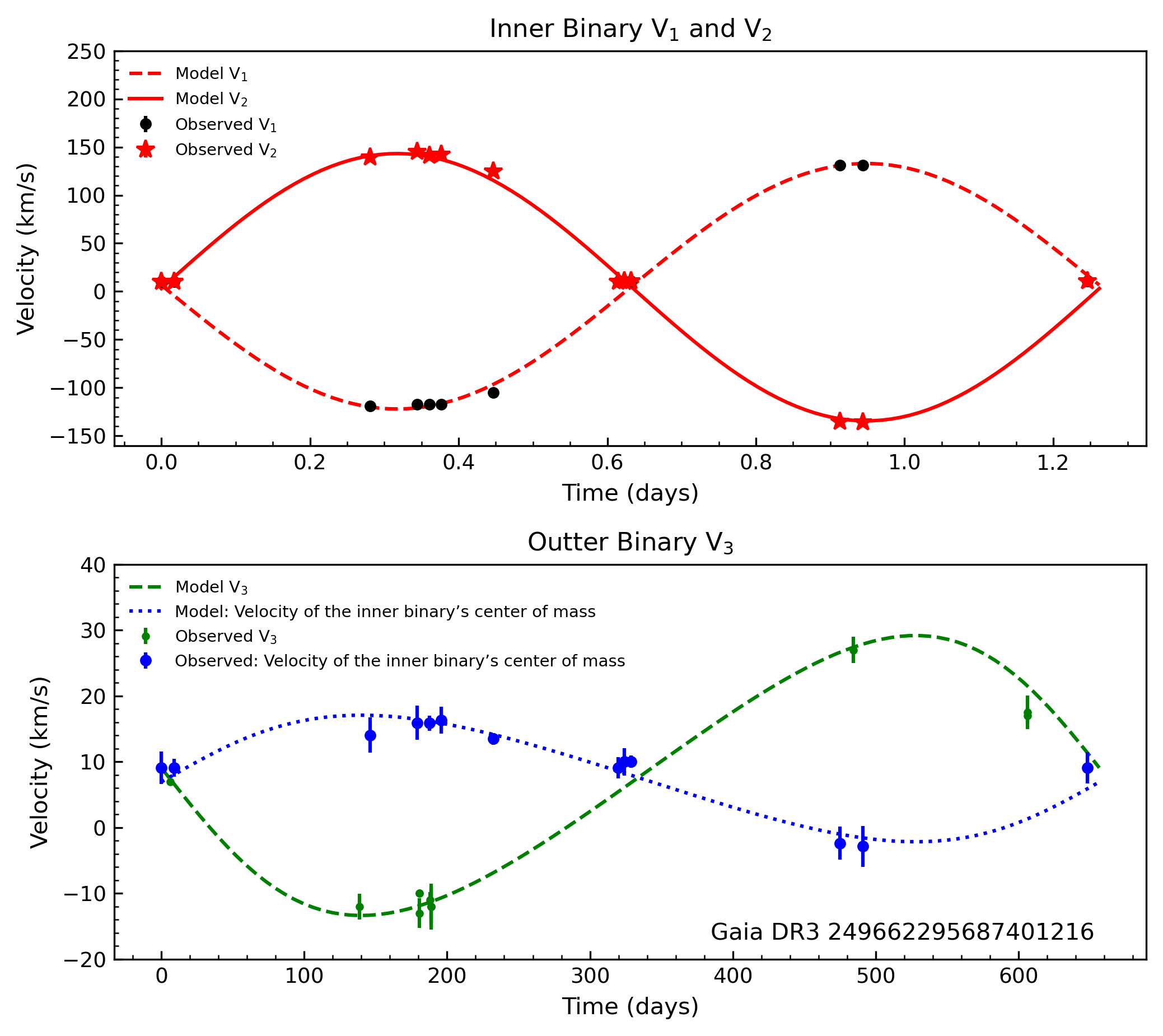}
\caption{RV fitting for the triple star system $Gaia$ DR3 249662295687401216. The black points and red stars represent the observed data after phase folding, and the red line indicates the fitted RV curve. The blue points show the velocity of the inner binary’s center of mass, while the blue dashed line represents the fitted curve for the inner binary’s center of mass velocity. The green points indicate the velocity of the tertiary star, and the green dashed line illustrates the fitted curve for the tertiary star’s velocity. $V_{1}$ and $V_{2}$ denote the RVs of the primary and secondary stars in the inner binary, respectively, and $V_{3}$ represents the RV curve of the tertiary star.
\label{figure 4}}
\end{figure}

Given that both the outer and inner periods of the triple are known—where the outer periods are obtained from the $Gaia$ NSS catalog and the inner periods are derived from $TESS$ light curves (the inner and outer periods are 1.26235 days and 656.47206 days. we folded the velocity-time diagrams into phase based on these period). Figure \ref{figure 4} illustrates this folding process. The black dots and red stars in figure \ref{figure 4} denote the velocity maps obtained after folding of the inner and outer periods, and the green points show the velocity of the inner binary’s center of mass, while the blue dashed line represents the fitted curve for the inner binary’s center of mass velocity. $V_{1}$,$V_{2}$,$V_{3}$ denote the apparent velocity profile for one period, respectively. $V_{1}$ and $V_{2}$ denote the RV curves of the primary and secondary stars in the inner binary, respectively, and $V_{3}$ represents the RV curve of the third star. 

With the RV curves, we can proceed to fit these curves using the Radvel package \citep{2018Fulton,2024lijiao} to derive various parameters. For the triple-star systems, we apply $Kepler$'s laws and classical Newtonian dynamics, treating the inner binary as a single entity due to the small ratio between the inner and outer periods. This approach allows us to calculate the orbital parameters of the center of mass of the inner binary with respect to the third star.

In the analysis, we define the likelihood function for the RV data by comparing the observed values with the model predictions. The model accounts for the contributions of each component in the triple-star system and includes an error term for each observation. Our prior distributions are as follows: for example, the RV amplitudes \(K_1\) and \(K_2\) of the inner binary are constrained within known ranges, the eccentricity \(e_2\) of the outer orbit is derived from the $Gaia$ NSS catalog data and its errors, and the mass ratios \(q_1\) of the inner binary and \(q_2\) of the third star are constrained between 0 and 1. We then use the Markov Chain Monte Carlo (MCMC; \cite{2013Foreman}) method for parameter sampling, with 100 walkers and 50,000 iterations, where the first 5,000 iterations are designated as the burn-in phase. The initial parameter distributions are drawn from the priors. By analyzing the MCMC chains, we extract the posterior distributions of parameters such as the RV amplitudes \(K_b\), \(K_c\), and \(K_a\), as well as the systemic velocity \(\gamma\). We assessed the fit quality by examining the residuals between the observed data and the model predictions and calculate the logarithmic probability to ensure that the chains have thoroughly explored the parameter space and meet the constraints.

We used MCMC to determine some orbital parameters of the triple-star systems. The parameters are as follows: for the inner binary: \(q_1 = 0.919 \pm 0.006\), \(K_1 = 138.798^{+0.727}_{-0.749}\), \(\sqrt{e_1}\cos\omega_1 = 0\), \(\sqrt{e_1}\sin\omega_1 = 0.0014 \pm 0.001\); Outer binary: \(\sqrt{e_2}\cos\omega_2 = -0.008^{+0.015}_{-0.008}\), \(\sqrt{e_2}\sin\omega_2 = 0.144^{+0.085}_{-0.068}\), \(q_2 = 0.0.427^{+0.077}_{-0.068}\), \(K_3 = 21.275^{+1.752}_{-1.484}\); and the system velocity \(\gamma = 7.74 ^{+1.038}_{-0.988} \ \text{km/s}\). See figures \ref{figure m1} in the appendix for more details.

\subsubsection{Fitting of light curve}
\label{sec:4.2.2}

\begin{figure}[ht!]
\centering
\includegraphics[width=11cm]{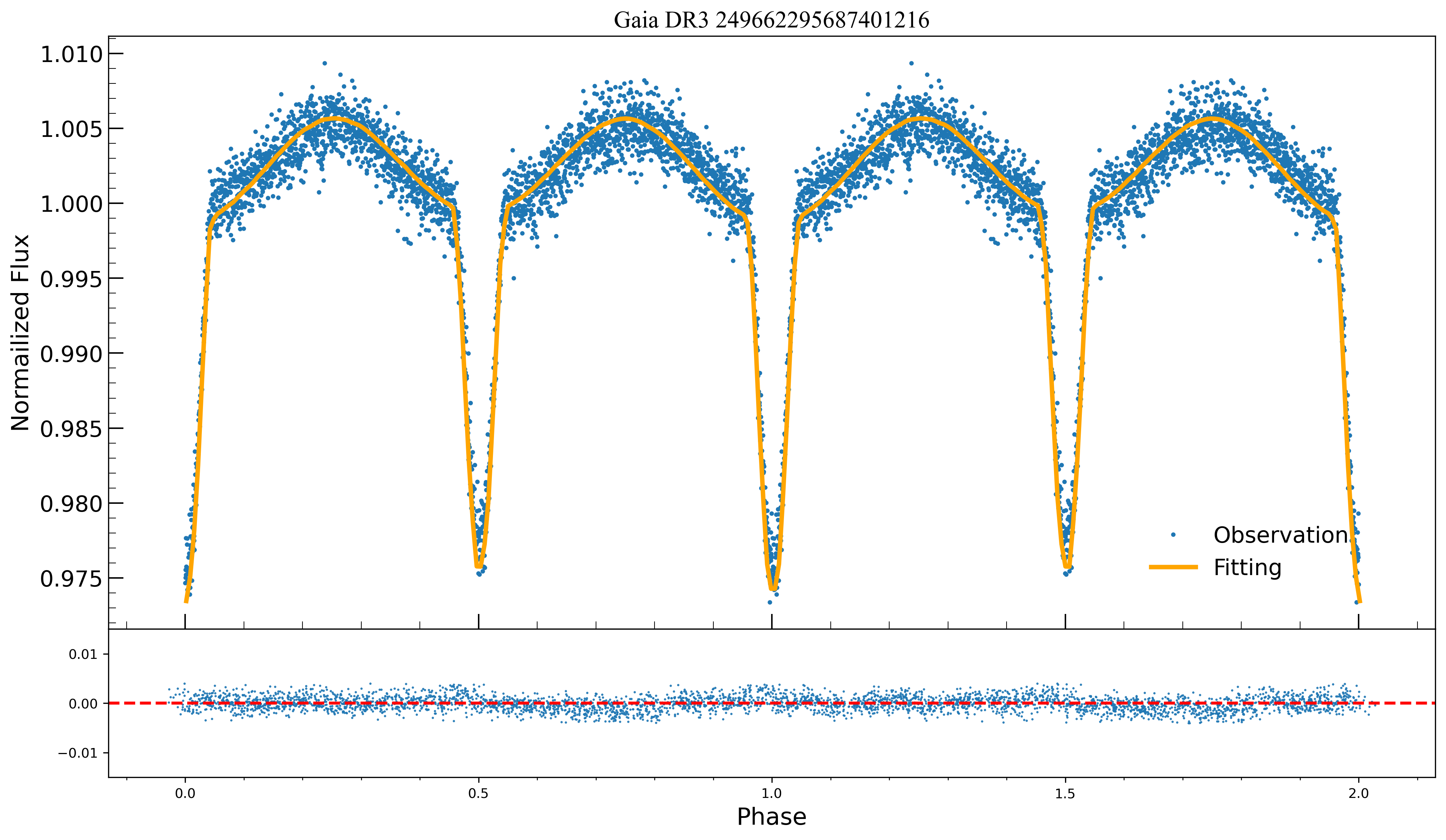}
\caption{Light curve fitting for the triple star system $Gaia$ DR3 249662295687401216. The blue dots indicate phase-folded observations, and the solid yellow line indicates the results of the fit. The horizontal axis indicates the phase, and the fitting residuals are shown at the bottom of the plot.
\label{figure 7}}
\end{figure}

We identified the light curve of the triple-star system from $TESS$ data \citep{Ricker2015}. Lomb–Scargle periodogram analysis \citep{2017Lomb} of the light curve yielded inner periods of 1.2623 days. This is consistent with the criterion of an outer-to-inner period ratio greater than 5 for triple-star systems \citep{1995ApJ...455..640E}. Subsequently, we folded the light curves to phase, as shown in figure \ref{figure 7}. Observational data are indicated by blue points.

We then employed the PHOEBE software (Physics of Eclipsing Binaries, PHOEBE 2.26) \citep{2005Pr, 2016Pr,2020Jones} to perform parameter fitting for the light curves. PHOEBE is based on the WD code by \citep{1971Wilson}. The eclipsing light curves constrain the inclination angle and relative radius of the secondary star with respect to the semi-major axis \citep{2024Xiong}. Given that the light curves visually resemble those of detached binaries, we applied the detached binary model. The effective temperature and surface gravity (\(\log g\)) of the primary star were estimated from the spectrum at the secondary eclipse minimum, where the contribution of the secondary star to the total flux is minimal. \cite{2018El-Badry} found that typical errors in effective temperature (T$_{\text{eff}}$) and surface gravity (\(\log g\)) for unresolved binaries treated as single stars are less than 200 K and 0.1 dex, respectively, for LAMOST spectra. These systematic errors are comparable to measurement uncertainties and do not significantly affect the final results for mass and radius estimates \citep{2023Xiong}.

We employed Markov Chain Monte Carlo (MCMC; 
\cite{2013Foreman}) sampling in PHOEBE to solve for 
all orbital parameters. The following prior 
information was added to constrain the MCMC 
sampling:\\
\\
1. Orbital inclination (\(i\)): \(0^\circ < i < 90^\circ\)  \\
2. Temperature ratio of the secondary to the primary star: \(0 < T_{\text{2}} / T_{\text{1}} < 1\)   \\
3. Effective radius of the primary star (\(R_1/a\)): \(0 < R_1/a < 1\)   \\
4. Effective radius of the secondary star (\(R_2/a\)): \(0 < R_2/a < 1\)  \\
4. Third light fraction : $0 < L_{3} <1$\\

We used the temperature obtained from LAMOST as the fixed parameter for the primary star and performed a search with 80 walkers for 20,000 steps. The peak of the probability distribution and the standard deviation of randomly sampled points were calculated as the best-fit parameters and their uncertainties. With these parameters, we can determine the inclination, temperature ratio of the primary and secondary stars, \(R_1/a\), \(R_2/a\)  and $L_{3}$. We adopted the third-light model ‘\textit{with third light}’ in Phoebe. Figures \ref{figure 7} and \ref{figure 6} present the results obtained by fitting the light curves of this triple system using Phoebe. From figures \ref{figure 7}, it is evident that the blue points represent the phase-folded observational data, while the yellow solid lines indicate the fitting results. The horizontal axis denotes the phase, and the fitting residuals are shown at the bottom of the figures.

Utilizing the light curves from $TESS$, we ultimately derived the following parameters: the inclination angle $i$ is $69.854^{+0.631}_{-0.461}$, $T_2/T_1=0.980^{+0.007}_{-0.009}$, $R_1/a=0.216^{+0.009}_{-0.02}$, $R_2/a=0.206^{+0.014}_{-0.017}$, and $L_3=0.443^{+0.044}_{-0.046}$.


However, \(R_1/a\) and \(R_2/a\) are only the relative radii of the two stars, and accurate values for the stellar radii cannot be obtained solely from the light curve. Therefore, we can use the RV curves and the light curves to determine the parameters of the components in the triple-star systems. Using the data in the \ref{sec:4.1} section, we obtained the eccentricities ($e_{1}$, $e_{2}$) and velocity amplitudes ($K_{1}$, $K_{2}$) for the inner and outer binary stars. Combining these with the inclination obtained from the light curve, we can determine the semi-major axis \(a\) of the inner binary star using the following formula \citep{2023Xiong,2024lijiao}:

\begin{equation}
K_1 = \frac{2\pi a_1 \sin i}{P_{1}(1 - e_{1}^2)^{1/2}},
\end{equation}

\begin{equation}
K_2 = \frac{2\pi a_2 \sin i}{P_{1}(1 - e_{1}^2)^{1/2}}.
\end{equation}

$P_{1}$ represents the inner period, $e$ represents the eccentricity of the inner binary, $i$ is the inclination of the inner binary, which can be obtained from figs. \ref{figure 6}, and $K_{1}$ $K_{2}$ are the apparent velocity amplitudes of the inner binary, which can be obtained from Figs. \ref{figure m1}. With the semi-major axis \(a\)=$a_{1} +a_{2}$, we can determine the radii \(R_1\) and \(R_2\) of the inner binary stars. Next, the masses \(M_1\) and \(M_2\) of the inner binary stars can be calculated using the following equations:

\begin{equation}
f_{\text{binary}} = \frac{M_{2}^3 \sin^3 i_1}{(M_{1} + M_{2})^2} = \frac{P_1 K_{1}^3 (1 - e_1^2)^{3/2}}{2\pi G}
\end{equation}

$M_{1}$ represents the mass of the primary star in the inner binary and $M_{2}$ represents the mass of the secondary star in the inner binary. The mass of the third star can be derived from the mass ratio \(q_2\), and then the Roche lobe radius can be calculated using the following equation to determine whether the inner binary stars overflow their Roche lobes, and The $q$ below represents the inner binary mass ratio $M_{2}/M_{1}$:

\begin{equation}
\frac{R_{\text{OL}}}{\text{a}} \equiv x_{\text{OL}}(q) \approx \frac{0.49 q^{2/3} + 0.27q - 0.12q^{4/3}}{0.6q^{2/3} + \ln(1 + q^{1/3})}, \quad q \leq 1,
\end{equation}







We obtained the specific parameters for this triple, as detailed in table \ref{tab2}. This table presents the parameters of the hierarchical triple-star systems analyzed in this study. This system $Gaia$ DR3 249662295687401216 is modeled using photometric and spectroscopic data. The parameters include the masses, temperatures, surface gravities, radii, Roche lobe radii, semi-major axes, and eccentricities for the inner and outer orbits, as well as the orbital periods. 

\begin{table}[ht]
\centering
\caption{Parameters of the triple star system $Gaia$ DR3 249662295687401216.\label{tab2}}
\begin{tabular}{llcc}
\hline
Parameter & Symbol & Triple : $Gaia$ DR3 249662295687401216  \\
\hline
BJD & $T_0$ & $2450000.216 \pm 0.006$ \\ 
Primary mass ($M_\odot$) & $M_1$ & $1.844 ^{+0.046}_{-0.045}$\\
Primary temperature (K) & $T_1$ & $7862.50 \pm 101.6$ \\
Primary surface gravity (log g) & $\log g_1$ & $4.336 ^{+0.085}_{-0.078}$ \\
Secondary mass ($M_\odot$) & $M_2$ & $1.694 \pm 0.034$  \\
Secondary temperature (K) & $T_2$ & $7711.85 \pm 122.70$ \\
Secondary surface gravity (log g) & $\log g_2$ & $4.340 ^{+0.074}_{-0.069}$ \\
Primary radius ($R_\odot$) & $R_1$ & $1.528 \pm 0.14$ \\ 
Secondary radius ($R_\odot$) & $R_2$ & $1.458 ^{+0.12}_{-0.12}$ \\ 
Roche lobe radius ($R_\odot$) & $R_L$ & $2.658 \pm 0.021$ \\ 
Semi-major axis inner ($R_\odot$) & $a_\text{inner}$ & $7.077 ^{+0.036}_{-0.035}$ \\ 
Inner eccentricity & $e_\text{inner}$ & $0.001 \pm 0.0010$ \\ 
Tertiary mass ($M_\odot$) & $M_3$ & $1.508 \pm ^{+0.27}_{-0.24}$ \\ 
Semi-major axis outer ($R_\odot$) & $a_\text{outer}$ & $545.127 ^{+10.31}_{-9.25}$ \\ 
Outer eccentricity & $e_\text{outer}$ & $0.144 \pm 0.067$ \\ 
Inner period (days) & $P_\text{in}$ & $1.262 \pm 0.0$ \\ 
Outer period (days) & $P_\text{out}$ & $656.472 \pm 14.85$ \\ 
Inner inclination (deg) & $i_\text{in}$ & $69.854 \pm 0.54$ \\ 
Outer inclination (deg) & $i_\text{out}$ & $89.90 \pm 27.29$ \\
\hline
\end{tabular}
\end{table}


\subsection{Determination of Orbital Parameters for Gaia DR3 2077667962475652864}
\label{sec:4.2}

Similarly, for the $Gaia$ DR3 2077667962475652864 triple system, we applied the same method by selecting the most recent spectroscopic data from the LAMOST middle-resolution spectra for analysis \ref{sec:4.1}. Additionally, we incorporated RV data from HIDES \citep{2017miniak,2022pan}. The details of the LAMOST data are provided in Table \ref{tab5} in the Appendix. Using a combination of RV fitting and light curve modeling, we derived the orbital parameters of the triple system. A detailed comparison of the RV and light curve fits can be found in figures \ref{figure t1},\ref{figure t2} and \ref{figure t3} in the Appendix. We then compared our results with \cite{2022pan}'s work, as shown in table \ref{tab3}. The comparison results indicate that our findings are consistent with most aspects of previous studies. The masses values of the inner binary stars in our research are slightly higher than those reported earlier, which may be attributed to the fact that we did not apply a zero-point offset correction among the RVs \citep{2021ApJS}.



\begin{table}[ht]
\centering
\caption{Parameters of the triple $Gaia$ DR3 2077667962475652864.\label{tab3}}
\begin{tabular}{llcc}
\hline
Parameter & Symbol & This work  & \cite{2022pan}'s work \\ \hline
Semi-amplitude of the third star's RV (km/s) & $K_{3}$ & $30.001^{+0.049}_{-0.047}$ & $30.08\pm 0.08$  \\
RV semi-amplitude of the inner binary M$_{1}$(km/s) & $K_{1}$ & $91.971^{+0.16}_{-0.15}$ & -  \\
$M_{2}/M_{1}$ & $q_{\text{in}}$ & $0.937\pm 0.003$ & $0.939\pm 0.001$ \\ 
$M_{3}/(M_{1}+M_{2})$ & $q_{\text{out}}$ & $0.393\pm 0.003$ & $0.387\pm 0.005$ \\ 
Inner period (days) & $P_{\text{in}}$ & $3.421\pm 0.0006$ & 3.421 \\ \
Outer period (days) & $P_{\text{out}}$ & $422.559\pm 3.33$ (fixed) & $418.2\pm 0.20$ \\ 
Inner inclination incl (deg) & $i_{\text{in}}$ & $86.163\pm 0.75$ & $85.34\pm 0.01$ \\
Outter inclination incl (deg) & $i_{\text{out}}$ & $89.90\pm 6.46$ & $84.50\pm 1.90$ \\ 
Inner eccentricity & $e_{\text{inner}}$ & $0.008\pm 0.001$ & 0.00 \\ 
Outer eccentricity & $e_{\text{outer}}$ & $0.294\pm 0.0014$ & $0.300\pm 0.002$ \\ 
Primary temperature (K) & $T_{1}$ & $6105.2\pm 149.2$ (fixed) & 6103 \\ 
Secondary temperature (K) & $T_{2}$ & $5879.308\pm 155.1$ & $5950\pm 100$ \\
System velocity (km/s) & $\gamma$ & $4.582^{+0.023}_{-0.025}$ & $4.56 \pm 0.04 $  \\
Primary radius$(R_{\odot})$ & $R_{1}$ & $1.098^{+0.12}_{-0.12}$ & $1.127\pm 0.008$ \\ 
Secondary radius (R$_{\odot})$ & $R_{2}$ & $0.880^{+0.12}_{-0.12}$ & $0.963\pm 0.007$ \\ 
Semi-major axis inner$(R_{\odot})$ & $a_{\text{inner}}$ & $12.076^{+0.029}_{-0.023}$ & $12.02\pm 0.01$ \\ 
Primary mass$(M_{\odot})$ & $M_{1}$ & $1.186\pm 0.013$ & $1.0286\pm 0.0026$ \\ 
Secondary mass ($M_{\odot}$) & $M_{2}$ & $1.112\pm 0.010$ & $0.9667\pm 0.0024$ \\ 
Primary surface gravity (log g) & $\log g_1$ & $4.431^{+0.10}_{-0.10}$ & $4.35\pm 0.01$ \\ 
Secondary surface gravity (log g) & $\log g_2$ & $4.596^{+0.13}_{-0.12}$ & $4.45\pm 0.01$ \\ 
Semi-major axis outer$(R_{\odot})$ & $a_{\text{outer}}$ & $346.898^{+1.02}_{-0.93}$ & $329.73\pm 1.72$ \\ 
The third light & $L_{3}$ & $0.349\pm 0.11$ & $0.257\pm 0.002$ \\
Tertiary mass ($M_{\odot}$) & $M_{3}$ & $0.904\pm 0.01$ & - \\ 
Roche lobe radius$(R_{\odot})$ & R$_{\text{L}}$ & $4.562\pm 0.013$ & - \\ 
\hline
\end{tabular}
\end{table}

\section{Discussion of Radial Velocities (RV) for the 23 Triple Systems from LAMOST and the j03 Triple System}\label{sec:6}

\begin{figure}[ht!]
\centering
\includegraphics[width=11cm]{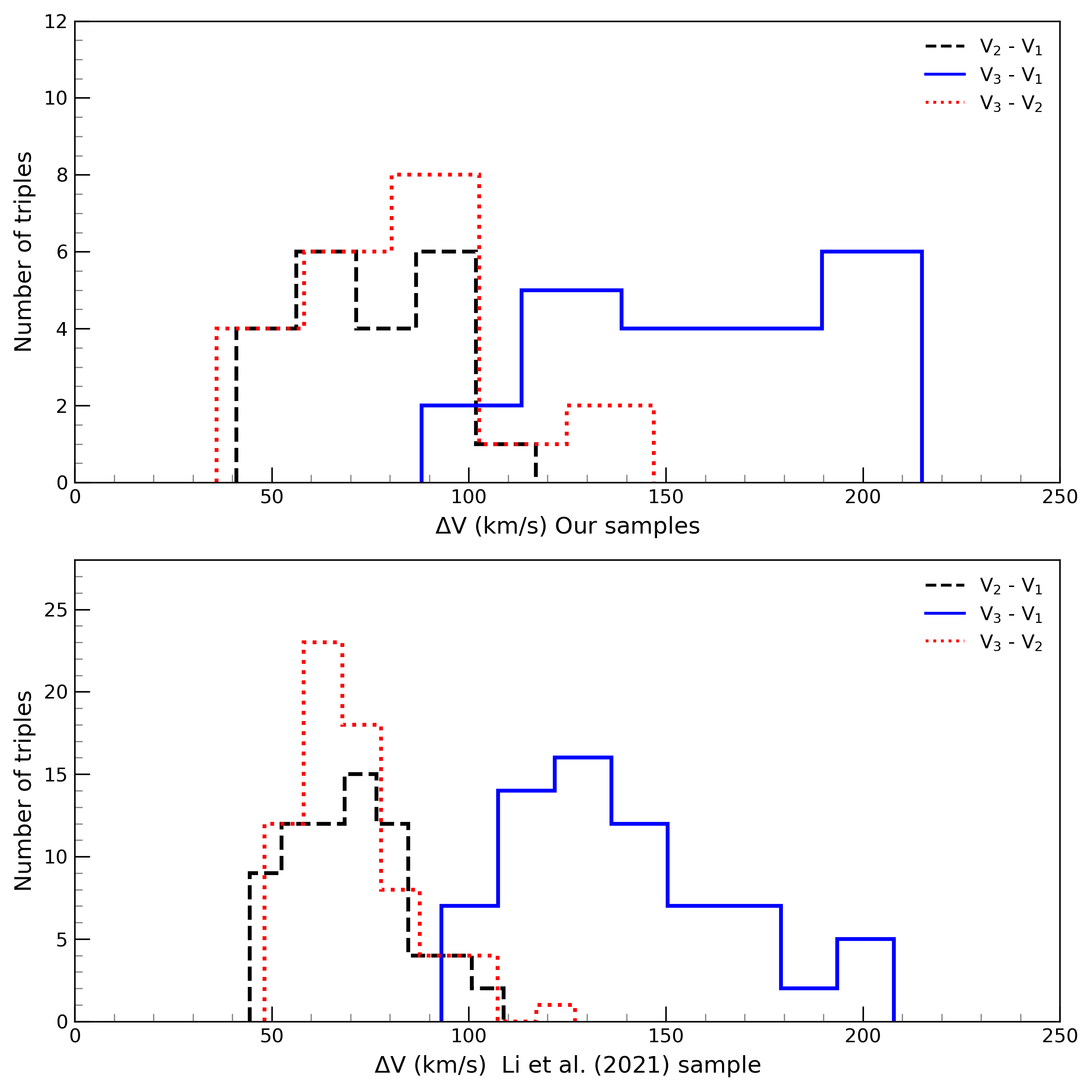}
\caption{Distribution of the number of $\Delta V$s between any two components in the triples. The three velocities, ordered from lowest to highest, are labeled as \( V_{1} \), \( V_{2} \), and \( V_{3} \). The upper panel shows the $\Delta V$ distribution between any two components of the triple star sample from Section \ref{sec:3}, while the lower panel shows the same for the 132 triple star candidates from \cite{li-2021}. The black dashed line denotes $\Delta V$ between $V_{2} - V_{1}$ ($\Delta V_{2-1}$), the red dash-dotted line shows $\Delta V$ between $V_{3} - V_{2}$ ($\Delta V_{3-2}$), and the blue solid line indicates $\Delta V$ between $V_{3} - V_{1}$ ($\Delta V_{3-1}$).
\label{figure 13}}
\end{figure}
In this section, we discussed the distribution of RVs for the 23 triple systems identified in the LAMOST dataset and discuss one triple system j03. 

We compared the RV differences ($\Delta V$) between components in our triple stars from Section \ref{sec:3} with those of 132 triple candidates identified by \cite{li-2021} using LAMOST-MRS spectra. Specifically, we examined the distribution of the RV differences ($\Delta V$) between any two components in these two samples. We designated the three velocities in ascending order as $V_{1}$, $V_{2}$, and $V_{3}$. The upper panel of figure \ref{figure 13} shows the $\Delta V$ distribution in our triples, while the lower panel illustrates the same for the \cite{li-2021}'s sample. The black dashed line represents the velocity difference between $V_{2} - V_{1}$ ($\Delta V_{2-1}$), the red dash-dotted line denotes the difference between $V_{3} - V_{2}$ ($\Delta V_{3-2}$), and the blue solid line indicates the difference between $V_{3} - V_{1}$ ($\Delta V_{3-1}$). The definitions of $\Delta V$ are detailed in figure \ref{figure 1}.  

The minimum $\Delta V$ in our triples is approximately 40 km/s, aligning with the resolution of LAMOST-MRS spectra, and the maximum $\Delta V$ is around 210 km/s. These results are consistent with those from \cite{li-2021}, confirming the reliability of our analysis. The $\Delta V$ results also show that components of the triple systems can be effectively resolved using the CCF method. Notably, we cannot detect velocity differences less than 45 km/s (spectral resolution constraints) in binary stars or below 90 km/s in the inner binaries of triple systems. According to figure 9 in \cite{He2023}, we established the relationship between the inner periods of triple systems and detection efficiency by analyzing various orbital period ratios ($P_{\text{out}}/P_{\text{in}}$). Due to the constraints of spectral resolution and RV, our detection range for inner periods is limited to between 0.2 days and 20 days, with maximum detection efficiency for periods shorter than 10 days. This observation suggests that the spectroscopic observations method has a distinct advantage in detecting short-period triple systems. 

Next, we discussed the triple system j03 ($Gaia$ DR3 66913361387671424). In this system, the primary star in the inner binary and the tertiary are relatively bright, while the secondary star is much fainter. In \cite{2024Kovalev}, they employed spectral disentangling to analyse the system. They precisely determined an inner period of 5.73 days, the RV semi-amplitude of one of the stars and the systemic velocity, among other parameters, and roughly estimated the mass ratio of the inner binary as 0.63. However, due to the lack of information about the outer orbit and the fact that they could only accurately obtain the RV of one star in the inner binary, they were unable to determine the inner binary’s mass ratio from the RV amplitude.

\begin{figure}[ht!]
\centering
\includegraphics[width=11cm]{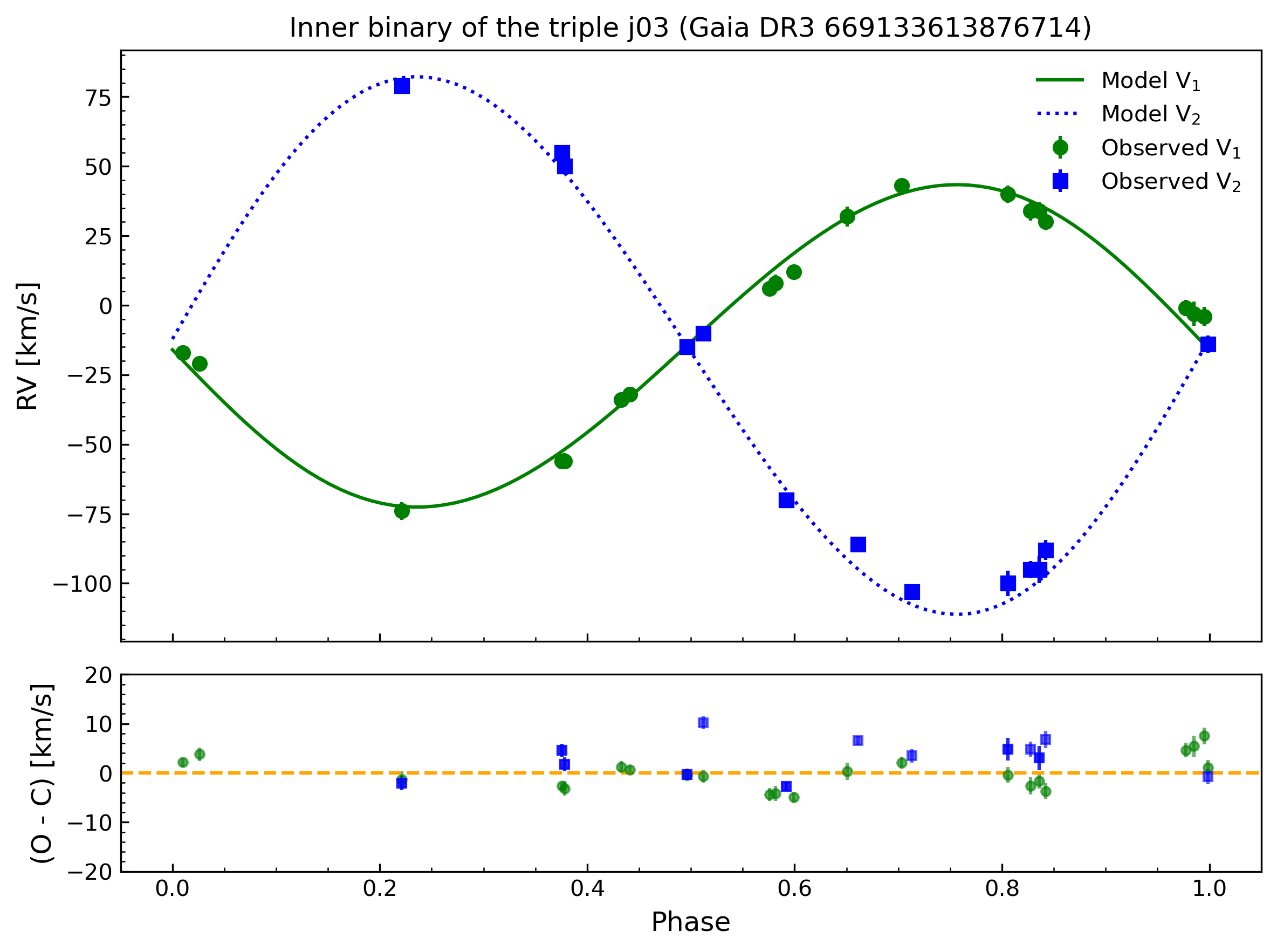}
\caption{RV fitting of inner binaries for the triple star system j03 $Gaia$ DR3 669133613876714. The green circles show the velocity of the primary star, while blue squares represent the velocity of the secondary star. $V_{1}$ and $V_{2}$ denote the RV curves of the primary and secondary stars in the inner binary.
\label{figure tt1}}
\end{figure}

In our approach, we employed template matching to directly obtain the RV curve of the secondary star in the inner binary. With both RV curves available, we can accurately determine the mass ratio. As shown in the figure \ref{figure tt1}, our RV curve fitting yields a precise mass ratio of 0.600 $^{+0.008}_{-0.009}$ (see figure \ref{figure tt2} in the Appendix). At present, it is unclear whether these fitting residuals are due to the influence of the third star. We obtained a comparable mass ratio to previous results, which suggests that our measurements are reliable.


\section{Summary and Conslusion}\label{sec:7}

In this study, we successfully identified and characterized new hierarchical triple-star systems by leveraging the synergy between $Gaia$ DR3 NSS and LAMOST DR10 data. By cross-matching these datasets, we discovered 23 triple-star systems, including 18 new systems. For two triples with extensive observations, we conducted cross-correlation function (CCF) spectral analysis, RV fitting, and light curve analysis. We determined the system parameters of these triples. Our approach highlights the complementary nature of astrometry and spectroscopy in the detection and characterization of multiple star systems. $Gaia$'s astrometry is highly sensitive to long-period and wide-orbit systems, whereas LAMOST's spectroscopy effectively reveals short-period systems. This dual approach enhances the overall efficiency of detecting and studying hierarchical triple-star systems \citep{Gaia2023}.

For the two triples, $Gaia$ DR3 249662295687401216, and $Gaia$ DR3 2077667962475652864, we utilized the Lomb-Scargle periodogram to determine their inner periods. We then phase-folded the light curves, revealing detailed periodic variations \citep{2024Moharana}. Subsequently, we modeled these triples using the PHOEBE software, obtaining key parameters such as orbital inclination, temperature ratio, relative radii, and the third light contribution. We refined these parameters precisely using MCMC sampling methods. The RV curves derived from our analysis were fitted using the \text{Radvel} package, yielding reliable orbital parameters. By combining RV data with light curves, we employed $Kepler$'s laws to calculate the semi-major axis and mass of the inner binary. This comprehensive approach allowed us to accurately determine the physical properties of the hierarchical triple-star systems, including the masses, radii, and orbital configurations of the constituent stars.


Then we analyzed the RV distributions of 23 triple systems from the LAMOST dataset. The RV differences ($\Delta V$) in our triples ranged from about 40 km/s to 210 km/s, confirming the reliability of our findings and limiting detectable inner periods to between 0.2 and 20 days. For the triple system j03, we utilized template matching to derive the RV curve of the secondary star in the inner binary. This differs from the spectral disentangling approach in \cite{2024Kovalev}, they estimated the mass ratio of the inner orbital as approximately 0.63. Our method, which includes both RV curves, yielded a precise mass ratio of 0.6 $^{+0.008}_{-0.009}$,  illustrating the reliability of our approach.



The study of multiple-star systems, especially triple star systems, is of significant importance in astronomy. They play a crucial role in stellar dynamics and evolution, providing valuable insights into stellar interactions and mass transfer \citep{Tokovinin2006,2010ragha}. Combining the strengths of astrometry and spectroscopy can provide new ways of revealing the complex structure and distribution properties of these systems, contributing to our understanding of star formation and evolution.

\section*{Acknowledgments}
We thank Hongwei Ge and Dengkai Jiang for their helpful insights. This work is supported by the National Natural Science Foundation of China under Grant Nos.12288102, 12125303, 12090040/3, 12373037, the National Key R\&D Program of China (Nos. 2021YFA1600403/1 and 2021YFA1600400), the Natural Science Foundation of Yunnan Province (Nos. 202201BC070003, 202001AW070007), the International Centre of Supernovae, Yunnan Key Laboratory (No. 202302AN360001), and the "Yunnan Revitalization Talent Support Program"-Science and Technology Champion Project (No. 202305AB350003). Guoshoujing Telescope (LAMOST) is a National Major Scientific Project built by the Chinese Academy of Sciences. Funding for the project has been provided by the National Development and Reform Commission. LAMOST is operated and managed by the National Astronomical Observatories, Chinese Academy of Sciences. This paper includes data collected by the $TESS$ mission. The TESS mission is funded by NASA’s Explorer Program. This work has also made use of data from the European Space Agency (ESA) mission $Gaia$ (\url{https://www.cosmos.esa.int/gaia}, \url{https://gea.esac.esa.int/archive/documentation/GDR3/index.html}).








\begin{thebibliography}{}
\expandafter\ifx\csname natexlab\endcsname\relax\def\natexlab#1{#1}\fi
\providecommand{\url}[1]{\href{#1}{#1}}
\providecommand{\dodoi}[1]{doi:~\href{http://doi.org/#1}{\nolinkurl{#1}}}
\providecommand{\doeprint}[1]{\href{http://ascl.net/#1}{\nolinkurl{http://ascl.net/#1}}}
\providecommand{\doarXiv}[1]{\href{https://arxiv.org/abs/#1}{\nolinkurl{https://arxiv.org/abs/#1}}}

\bibitem[{{Antonini} {et~al.}(2017){Antonini}, {Toonen}, \&
  {Hamers}}]{Fabio2017}
{Antonini}, F., {Toonen}, S., \& {Hamers}, A.~S. 2017, \apj, 841, 77,
  \dodoi{10.3847/1538-4357/aa6f5e}

\bibitem[{{Blanco-Cuaresma}(2019)}]{2019Blanco}
{Blanco-Cuaresma}, S. 2019, \mnras, 486, 2075, \dodoi{10.1093/mnras/stz549}

\bibitem[{{Blanco-Cuaresma} {et~al.}(2014){Blanco-Cuaresma}, {Soubiran},
  {Heiter}, \& {Jofr{\'e}}}]{2014ispec}
{Blanco-Cuaresma}, S., {Soubiran}, C., {Heiter}, U., \& {Jofr{\'e}}, P. 2014,
  \aap, 569, A111, \dodoi{10.1051/0004-6361/201423945}

\bibitem[{Cui {et~al.}(2012)Cui, Zhao, Chu, Li, Li, {Zhang}, {Su}, {Yao},
  {Wang}, {Xing}, {Li}, {Zhu}, {Wang}, {Gu}, {Luo}, {Xu}, {Zhang}, {Liu},
  {Zhang}, {Yang}, {Cao}, {Chen}, {Chen}, {Chen}, {Chen}, {Chu}, {Feng},
  {Gong}, {Hou}, {Hu}, {Hu}, {Hu}, {Jia}, {Jiang}, {Jiang}, {Jiang}, {Jin},
  {Li}, {Li}, {Li}, {Liu}, {Liu}, {Lu}, {Mao}, {Men}, {Qi}, {Qi}, {Shi},
  {Tang}, {Tao}, {Wang}, {Wang}, {Wang}, {Wang}, {Wang}, {Wang}, {Wang},
  {Wang}, {Wang}, {Wang}, {Wang}, {Wang}, {Xu}, {Xu}, {Yang}, {Yu}, {Yuan},
  {Yuan}, {Zhai}, {Zhang}, {Zhang}, {Zhang}, {Zhao}, {Zhou}, {Zhou}, {Zhu}, \&
  {Zou}}]{2012cui}
Cui, X., Zhao, Y., Chu, Y., {et~al.} 2012, Research in Astronomy and
  Astrophysics, 12, 1197, \dodoi{10.1088/1674-4527/12/9/003}

\bibitem[{{Czavalinga} {et~al.}(2023){Czavalinga}, {Mitnyan}, {Rappaport},
  {Borkovits}, {Gagliano}, {Omohundro}, {Kristiansen}, \& {P{\'a}l}}]{aa}
{Czavalinga}, D.~R., {Mitnyan}, T., {Rappaport}, S.~A., {et~al.} 2023, \aap,
  670, A75, \dodoi{10.1051/0004-6361/202245300}

\bibitem[{{Duch{\^e}ne} \& {Kraus}(2013)}]{2013ARA}
{Duch{\^e}ne}, G., \& {Kraus}, A. 2013, \araa, 51, 269,
  \dodoi{10.1146/annurev-astro-081710-102602}

\bibitem[{Duch{\^{e} }ne \& Kraus(2013)}]{Duch_ne_2013}
Duch{\^{e} }ne, G., \& Kraus, A. 2013, Annual Review of Astronomy and
  Astrophysics, 51, 269, \dodoi{10.1146/annurev-astro-081710-102602}

\bibitem[{{Eggleton} \& {Kiseleva}(1995)}]{1995ApJ...455..640E}
{Eggleton}, P., \& {Kiseleva}, L. 1995, \apj, 455, 640, \dodoi{10.1086/176611}

\bibitem[{{Eggleton} \& {Tokovinin}(2008)}]{2008eggleton}
{Eggleton}, P.~P., \& {Tokovinin}, A.~A. 2008, \mnras, 389, 869,
  \dodoi{10.1111/j.1365-2966.2008.13596.x}

\bibitem[{{Eggleton} \& {Tokovinin}(2010)}]{2010eggleton}
---. 2010, VizieR Online Data Catalog, J/MNRAS/389/869

\bibitem[{{El-Badry} {et~al.}(2018){El-Badry}, {Rix}, {Ting}, {Weisz},
  {Bergemann}, {Cargile}, {Conroy}, \& {Eilers}}]{2018El-Badry}
{El-Badry}, K., {Rix}, H.-W., {Ting}, Y.-S., {et~al.} 2018, \mnras, 473, 5043,
  \dodoi{10.1093/mnras/stx2758}

\bibitem[{{Fernandez} {et~al.}(2017){Fernandez}, {Covey}, {De Lee},
  {Chojnowski}, {Nidever}, {Ballantyne}, {Cottaar}, {Da Rio}, {Foster},
  {Majewski}, {Meyer}, {Reyna}, {Roberts}, {Skinner}, {Stassun}, {Tan},
  {Troup}, \& {Zasowski}}]{2017fer}
{Fernandez}, M.~A., {Covey}, K.~R., {De Lee}, N., {et~al.} 2017, \pasp, 129,
  084201, \dodoi{10.1088/1538-3873/aa77e0}

\bibitem[{{Foreman-Mackey} {et~al.}(2013){Foreman-Mackey}, {Hogg}, {Lang}, \&
  {Goodman}}]{2013Foreman}
{Foreman-Mackey}, D., {Hogg}, D.~W., {Lang}, D., \& {Goodman}, J. 2013, \pasp,
  125, 306, \dodoi{10.1086/670067}

\bibitem[{{Fulton} {et~al.}(2018){Fulton}, {Petigura}, {Blunt}, \&
  {Sinukoff}}]{2018Fulton}
{Fulton}, B.~J., {Petigura}, E.~A., {Blunt}, S., \& {Sinukoff}, E. 2018, \pasp,
  130, 044504, \dodoi{10.1088/1538-3873/aaaaa8}

\bibitem[{{Gaia Collaboration} {et~al.}(2023{\natexlab{a}}){Gaia
  Collaboration}, {Vallenari}, {Brown}, {Prusti}, {de Bruijne}, {Arenou},
  {Babusiaux}, {Biermann}, {Creevey}, {Ducourant}, {Evans}, {Eyer}, {Guerra},
  {Hutton}, {Jordi}, {Klioner}, {Lammers}, {Lindegren}, {Luri}, {Mignard},
  {Panem}, {Pourbaix}, {Randich}, {Sartoretti}, {Soubiran}, {Tanga}, {Walton},
  {Bailer-Jones}, {Bastian}, {Drimmel}, {Jansen}, {Katz}, {Lattanzi}, {van
  Leeuwen}, {Bakker}, {Cacciari}, {Casta{\~n}eda}, {De Angeli}, {Fabricius},
  {Fouesneau}, {Fr{\'e}mat}, {Galluccio}, {Guerrier}, {Heiter}, {Masana},
  {Messineo}, {Mowlavi}, {Nicolas}, {Nienartowicz}, {Pailler}, {Panuzzo},
  {Riclet}, {Roux}, {Seabroke}, {Sordo}, {Th{\'e}venin}, {Gracia-Abril},
  {Portell}, {Teyssier}, {Altmann}, {Andrae}, {Audard}, {Bellas-Velidis},
  {Benson}, {Berthier}, {Blomme}, {Burgess}, {Busonero}, {Busso},
  {C{\'a}novas}, {Carry}, {Cellino}, {Cheek}, {Clementini}, {Damerdji},
  {Davidson}, {de Teodoro}, {Nu{\~n}ez Campos}, {Delchambre}, {Dell'Oro},
  {Esquej}, {Fern{\'a}ndez-Hern{\'a}ndez}, {Fraile}, {Garabato},
  {Garc{\'\i}a-Lario}, {Gosset}, {Haigron}, {Halbwachs}, {Hambly}, {Harrison},
  {Hern{\'a}ndez}, {Hestroffer}, {Hodgkin}, {Holl}, {Jan{\ss}en}, {Jevardat de
  Fombelle}, {Jordan}, {Krone-Martins}, {Lanzafame}, {L{\"o}ffler}, {Marchal},
  {Marrese}, {Moitinho}, {Muinonen}, {Osborne}, {Pancino}, {Pauwels},
  {Recio-Blanco}, {Reyl{\'e}}, {Riello}, {Rimoldini}, {Roegiers}, {Rybizki},
  {Sarro}, {Siopis}, {Smith}, {Sozzetti}, {Utrilla}, {van Leeuwen}, {Abbas},
  {{\'A}brah{\'a}m}, {Abreu Aramburu}, {Aerts}, {Aguado}, {Ajaj},
  {Aldea-Montero}, {Altavilla}, {{\'A}lvarez}, {Alves}, {Anders}, {Anderson},
  {Anglada Varela}, {Antoja}, {Baines}, {Baker}, {Balaguer-N{\'u}{\~n}ez},
  {Balbinot}, {Balog}, {Barache}, {Barbato}, {Barros}, {Barstow},
  {Bartolom{\'e}}, {Bassilana}, {Bauchet}, {Becciani}, {Bellazzini},
  {Berihuete}, {Bernet}, {Bertone}, {Bianchi}, {Binnenfeld}, {Blanco-Cuaresma},
  {Blazere}, {Boch}, {Bombrun}, {Bossini}, {Bouquillon}, {Bragaglia},
  {Bramante}, {Breedt}, {Bressan}, {Brouillet}, {Brugaletta}, {Bucciarelli},
  {Burlacu}, {Butkevich}, {Buzzi}, {Caffau}, {Cancelliere}, {Cantat-Gaudin},
  {Carballo}, {Carlucci}, {Carnerero}, {Carrasco}, {Casamiquela}, {Castellani},
  {Castro-Ginard}, {Chaoul}, {Charlot}, {Chemin}, {Chiaramida}, {Chiavassa},
  {Chornay}, {Comoretto}, {Contursi}, {Cooper}, {Cornez}, {Cowell}, {Crifo},
  {Cropper}, {Crosta}, {Crowley}, {Dafonte}, {Dapergolas}, {David}, {David},
  {de Laverny}, {De Luise}, {De March}, {De Ridder}, {de Souza}, {de Torres},
  {del Peloso}, {del Pozo}, {Delbo}, {Delgado}, {Delisle}, {Demouchy},
  {Dharmawardena}, {Di Matteo}, {Diakite}, {Diener}, {Distefano}, {Dolding},
  {Edvardsson}, {Enke}, {Fabre}, {Fabrizio}, {Faigler}, {Fedorets}, {Fernique},
  {Fienga}, {Figueras}, {Fournier}, {Fouron}, {Fragkoudi}, {Gai},
  {Garcia-Gutierrez}, {Garcia-Reinaldos}, {Garc{\'\i}a-Torres}, {Garofalo},
  {Gavel}, {Gavras}, {Gerlach}, {Geyer}, {Giacobbe}, {Gilmore}, {Girona},
  {Giuffrida}, {Gomel}, {Gomez}, {Gonz{\'a}lez-N{\'u}{\~n}ez},
  {Gonz{\'a}lez-Santamar{\'\i}a}, {Gonz{\'a}lez-Vidal}, {Granvik}, {Guillout},
  {Guiraud}, {Guti{\'e}rrez-S{\'a}nchez}, {Guy}, {Hatzidimitriou}, {Hauser},
  {Haywood}, {Helmer}, {Helmi}, {Sarmiento}, {Hidalgo}, {Hilger},
  {H{\l}adczuk}, {Hobbs}, {Holland}, {Huckle}, {Jardine}, {Jasniewicz},
  {Jean-Antoine Piccolo}, {Jim{\'e}nez-Arranz}, {Jorissen}, {Juaristi
  Campillo}, {Julbe}, {Karbevska}, {Kervella}, {Khanna}, {Kontizas},
  {Kordopatis}, {Korn}, {K{\'o}sp{\'a}l}, {Kostrzewa-Rutkowska},
  {Kruszy{\'n}ska}, {Kun}, {Laizeau}, {Lambert}, {Lanza}, {Lasne}, {Le
  Campion}, {Lebreton}, {Lebzelter}, {Leccia}, {Leclerc}, {Lecoeur-Taibi},
  {Liao}, {Licata}, {Lindstr{\o}m}, {Lister}, {Livanou}, {Lobel}, {Lorca},
  {Loup}, {Madrero Pardo}, {Magdaleno Romeo}, {Managau}, {Mann}, {Manteiga},
  {Marchant}, {Marconi}, {Marcos}, {Marcos Santos}, {Mar{\'\i}n Pina},
  {Marinoni}, {Marocco}, {Marshall}, {Martin Polo}, {Mart{\'\i}n-Fleitas},
  {Marton}, {Mary}, {Masip}, {Massari}, {Mastrobuono-Battisti}, {Mazeh},
  {McMillan}, {Messina}, {Michalik}, {Millar}, {Mints}, {Molina}, {Molinaro},
  {Moln{\'a}r}, {Monari}, {Mongui{\'o}}, {Montegriffo}, {Montero}, {Mor},
  {Mora}, {Morbidelli}, {Morel}, {Morris}, {Muraveva}, {Murphy}, {Musella},
  {Nagy}, {Noval}, {Oca{\~n}a}, {Ogden}, {Ordenovic}, {Osinde}, {Pagani},
  {Pagano}, {Palaversa}, {Palicio}, {Pallas-Quintela}, {Panahi},
  {Payne-Wardenaar}, {Pe{\~n}alosa Esteller}, {Penttil{\"a}}, {Pichon},
  {Piersimoni}, {Pineau}, {Plachy}, {Plum}, {Poggio}, {Pr{\v{s}}a}, {Pulone},
  {Racero}, {Ragaini}, {Rainer}, {Raiteri}, {Rambaux}, {Ramos}, {Ramos-Lerate},
  {Re Fiorentin}, {Regibo}, {Richards}, {Rios Diaz}, {Ripepi}, {Riva}, {Rix},
  {Rixon}, {Robichon}, {Robin}, {Robin}, {Roelens}, {Rogues}, {Rohrbasser},
  {Romero-G{\'o}mez}, {Rowell}, {Royer}, {Ruz Mieres}, {Rybicki}, {Sadowski},
  {S{\'a}ez N{\'u}{\~n}ez}, {Sagrist{\`a} Sell{\'e}s}, {Sahlmann}, {Salguero},
  {Samaras}, {Sanchez Gimenez}, {Sanna}, {Santove{\~n}a}, {Sarasso},
  {Schultheis}, {Sciacca}, {Segol}, {Segovia}, {S{\'e}gransan}, {Semeux},
  {Shahaf}, {Siddiqui}, {Siebert}, {Siltala}, {Silvelo}, {Slezak}, {Slezak},
  {Smart}, {Snaith}, {Solano}, {Solitro}, {Souami}, {Souchay}, {Spagna},
  {Spina}, {Spoto}, {Steele}, {Steidelm{\"u}ller}, {Stephenson}, {S{\"u}veges},
  {Surdej}, {Szabados}, {Szegedi-Elek}, {Taris}, {Taylor}, {Teixeira},
  {Tolomei}, {Tonello}, {Torra}, {Torra}, {Torralba Elipe}, {Trabucchi},
  {Tsounis}, {Turon}, {Ulla}, {Unger}, {Vaillant}, {van Dillen}, {van Reeven},
  {Vanel}, {Vecchiato}, {Viala}, {Vicente}, {Voutsinas}, {Weiler}, {Wevers},
  {Wyrzykowski}, {Yoldas}, {Yvard}, {Zhao}, {Zorec}, {Zucker}, \&
  {Zwitter}}]{GaiaCollaboration2023}
{Gaia Collaboration}, {Vallenari}, A., {Brown}, A.~G.~A., {et~al.}
  2023{\natexlab{a}}, \aap, 674, A1, \dodoi{10.1051/0004-6361/202243940}

\bibitem[{{Gaia Collaboration} {et~al.}(2023{\natexlab{b}}){Gaia
  Collaboration}, {Arenou}, {Babusiaux}, {Barstow}, {Faigler}, {Jorissen},
  {Kervella}, {Mazeh}, {Mowlavi}, {Panuzzo}, {Sahlmann}, {Shahaf}, {Sozzetti},
  {Bauchet}, {Damerdji}, {Gavras}, {Giacobbe}, {Gosset}, {Halbwachs}, {Holl},
  {Lattanzi}, {Leclerc}, {Morel}, {Pourbaix}, {Re Fiorentin}, {Sadowski},
  {S{\'e}gransan}, {Siopis}, {Teyssier}, {Zwitter}, {Planquart}, {Brown},
  {Vallenari}, {Prusti}, {de Bruijne}, {Biermann}, {Creevey}, {Ducourant},
  {Evans}, {Eyer}, {Guerra}, {Hutton}, {Jordi}, {Klioner}, {Lammers},
  {Lindegren}, {Luri}, {Mignard}, {Panem}, {Randich}, {Sartoretti}, {Soubiran},
  {Tanga}, {Walton}, {Bailer-Jones}, {Bastian}, {Drimmel}, {Jansen}, {Katz},
  {van Leeuwen}, {Bakker}, {Cacciari}, {Casta{\~n}eda}, {De Angeli},
  {Fabricius}, {Fouesneau}, {Fr{\'e}mat}, {Galluccio}, {Guerrier}, {Heiter},
  {Masana}, {Messineo}, {Nicolas}, {Nienartowicz}, {Pailler}, {Riclet}, {Roux},
  {Seabroke}, {Sordo}, {Th{\'e}venin}, {Gracia-Abril}, {Portell}, {Altmann},
  {Andrae}, {Audard}, {Bellas-Velidis}, {Benson}, {Berthier}, {Blomme},
  {Burgess}, {Busonero}, {Busso}, {C{\'a}novas}, {Carry}, {Cellino}, {Cheek},
  {Clementini}, {Davidson}, {de Teodoro}, {Nu{\~n}ez Campos}, {Delchambre},
  {Dell'Oro}, {Esquej}, {Fern{\'a}ndez-Hern{\'a}ndez}, {Fraile}, {Garabato},
  {Garc{\'\i}a-Lario}, {Haigron}, {Hambly}, {Harrison}, {Hern{\'a}ndez},
  {Hestroffer}, {Hodgkin}, {Jan{\ss}en}, {Jevardat de Fombelle}, {Jordan},
  {Krone-Martins}, {Lanzafame}, {L{\"o}ffler}, {Marchal}, {Marrese},
  {Moitinho}, {Muinonen}, {Osborne}, {Pancino}, {Pauwels}, {Recio-Blanco},
  {Reyl{\'e}}, {Riello}, {Rimoldini}, {Roegiers}, {Rybizki}, {Sarro}, {Smith},
  {Utrilla}, {van Leeuwen}, {Abbas}, {{\'A}brah{\'a}m}, {Abreu Aramburu},
  {Aerts}, {Aguado}, {Ajaj}, {Aldea-Montero}, {Altavilla}, {{\'A}lvarez},
  {Alves}, {Anders}, {Anderson}, {Anglada Varela}, {Antoja}, {Baines}, {Baker},
  {Balaguer-N{\'u}{\~n}ez}, {Balbinot}, {Balog}, {Barache}, {Barbato},
  {Barros}, {Bartolom{\'e}}, {Bassilana}, {Becciani}, {Bellazzini},
  {Berihuete}, {Bernet}, {Bertone}, {Bianchi}, {Binnenfeld}, {Blanco-Cuaresma},
  {Blazere}, {Boch}, {Bombrun}, {Bossini}, {Bouquillon}, {Bragaglia},
  {Bramante}, {Breedt}, {Bressan}, {Brouillet}, {Brugaletta}, {Bucciarelli},
  {Burlacu}, {Butkevich}, {Buzzi}, {Caffau}, {Cancelliere}, {Cantat-Gaudin},
  {Carballo}, {Carlucci}, {Carnerero}, {Carrasco}, {Casamiquela}, {Castellani},
  {Castro-Ginard}, {Chaoul}, {Charlot}, {Chemin}, {Chiaramida}, {Chiavassa},
  {Chornay}, {Comoretto}, {Contursi}, {Cooper}, {Cornez}, {Cowell}, {Crifo},
  {Cropper}, {Crosta}, {Crowley}, {Dafonte}, {Dapergolas}, {David}, {de
  Laverny}, {De Luise}, {De March}, {De Ridder}, {de Souza}, {de Torres}, {del
  Peloso}, {del Pozo}, {Delbo}, {Delgado}, {Delisle}, {Demouchy},
  {Dharmawardena}, {Diakite}, {Diener}, {Distefano}, {Dolding}, {Enke},
  {Fabre}, {Fabrizio}, {Fedorets}, {Fernique}, {Figueras}, {Fournier},
  {Fouron}, {Fragkoudi}, {Gai}, {Garcia-Gutierrez}, {Garcia-Reinaldos},
  {Garc{\'\i}a-Torres}, {Garofalo}, {Gavel}, {Gerlach}, {Geyer}, {Gilmore},
  {Girona}, {Giuffrida}, {Gomel}, {Gomez}, {Gonz{\'a}lez-N{\'u}{\~n}ez},
  {Gonz{\'a}lez-Santamar{\'\i}a}, {Gonz{\'a}lez-Vidal}, {Granvik}, {Guillout},
  {Guiraud}, {Guti{\'e}rrez-S{\'a}nchez}, {Guy}, {Hatzidimitriou}, {Hauser},
  {Haywood}, {Helmer}, {Helmi}, {Sarmiento}, {Hidalgo}, {Hilger},
  {H{\l}adczuk}, {Hobbs}, {Holland}, {Huckle}, {Jardine}, {Jasniewicz},
  {Jean-Antoine Piccolo}, {Jim{\'e}nez-Arranz}, {Juaristi Campillo}, {Julbe},
  {Karbevska}, {Khanna}, {Kordopatis}, {Korn}, {K{\'o}sp{\'a}l},
  {Kostrzewa-Rutkowska}, {Kruszy{\'n}ska}, {Kun}, {Laizeau}, {Lambert},
  {Lanza}, {Lasne}, {Le Campion}, {Lebreton}, {Lebzelter}, {Leccia},
  {Lecoeur-Taibi}, {Liao}, {Licata}, {Lindstr{\o}m}, {Lister}, {Livanou},
  {Lobel}, {Lorca}, {Loup}, {Madrero Pardo}, {Magdaleno Romeo}, {Managau},
  {Mann}, {Manteiga}, {Marchant}, {Marconi}, {Marcos}, {Marcos Santos},
  {Mar{\'\i}n Pina}, {Marinoni}, {Marocco}, {Marshall}, {Martin Polo},
  {Mart{\'\i}n-Fleitas}, {Marton}, {Mary}, {Masip}, {Massari},
  {Mastrobuono-Battisti}, {McMillan}, {Messina}, {Michalik}, {Millar}, {Mints},
  {Molina}, {Molinaro}, {Moln{\'a}r}, {Monari}, {Mongui{\'o}}, {Montegriffo},
  {Montero}, {Mor}, {Mora}, {Morbidelli}, {Morris}, {Muraveva}, {Murphy},
  {Musella}, {Nagy}, {Noval}, {Oca{\~n}a}, {Ogden}, {Ordenovic}, {Osinde},
  {Pagani}, {Pagano}, {Palaversa}, {Palicio}, {Pallas-Quintela}, {Panahi},
  {Payne-Wardenaar}, {Pe{\~n}alosa Esteller}, {Penttil{\"a}}, {Pichon},
  {Piersimoni}, {Pineau}, {Plachy}, {Plum}, {Poggio}, {Pr{\v{s}}a}, {Pulone},
  {Racero}, {Ragaini}, {Rainer}, {Raiteri}, {Ramos}, {Ramos-Lerate}, {Regibo},
  {Richards}, {Rios Diaz}, {Ripepi}, {Riva}, {Rix}, {Rixon}, {Robichon},
  {Robin}, {Robin}, {Roelens}, {Rogues}, {Rohrbasser}, {Romero-G{\'o}mez},
  {Rowell}, {Royer}, {Ruz Mieres}, {Rybicki}, {S{\'a}ez N{\'u}{\~n}ez},
  {Sagrist{\`a} Sell{\'e}s}, {Salguero}, {Samaras}, {Sanchez Gimenez}, {Sanna},
  {Santove{\~n}a}, {Sarasso}, {Schultheis}, {Sciacca}, {Segol}, {Segovia},
  {Semeux}, {Siddiqui}, {Siebert}, {Siltala}, {Silvelo}, {Slezak}, {Slezak},
  {Smart}, {Snaith}, {Solano}, {Solitro}, {Souami}, {Souchay}, {Spagna},
  {Spina}, {Spoto}, {Steele}, {Steidelm{\"u}ller}, {Stephenson}, {S{\"u}veges},
  {Surdej}, {Szabados}, {Szegedi-Elek}, {Taris}, {Taylor}, {Teixeira},
  {Tolomei}, {Tonello}, {Torra}, {Torra}, {Torralba Elipe}, {Trabucchi},
  {Tsounis}, {Turon}, {Ulla}, {Unger}, {Vaillant}, {van Dillen}, {van Reeven},
  {Vanel}, {Vecchiato}, {Viala}, {Vicente}, {Voutsinas}, {Weiler}, {Wevers},
  {Wyrzykowski}, {Yoldas}, {Yvard}, {Zhao}, {Zorec}, \& {Zucker}}]{Gaia2023}
{Gaia Collaboration}, {Arenou}, F., {Babusiaux}, C., {et~al.}
  2023{\natexlab{b}}, \aap, 674, A34, \dodoi{10.1051/0004-6361/202243782}

\bibitem[{{Halbwachs} {et~al.}(2023){Halbwachs}, {Pourbaix}, {Arenou},
  {Galluccio}, {Guillout}, {Bauchet}, {Marchal}, {Sadowski}, \&
  {Teyssier}}]{2023Halbwachs}
{Halbwachs}, J.-L., {Pourbaix}, D., {Arenou}, F., {et~al.} 2023, \aap, 674, A9,
  \dodoi{10.1051/0004-6361/202243969}

\bibitem[{{He} {et~al.}(2023){He}, {Li}, {Chen}, {Yang}, {Xiao}, \&
  {Han}}]{He2023}
{He}, T., {Li}, J., {Chen}, X., {et~al.} 2023, \apj, 958, 14,
  \dodoi{10.3847/1538-4357/acf8c4}

\bibitem[{{He{\l}miniak} {et~al.}(2017){He{\l}miniak}, {Ukita}, {Kambe},
  {Koz{\l}owski}, {Sybilski}, {Maehara}, {Ratajczak}, {Konacki}, \&
  {Paw{\l}aszek}}]{2017miniak}
{He{\l}miniak}, K.~G., {Ukita}, N., {Kambe}, E., {et~al.} 2017, \mnras, 468,
  1726, \dodoi{10.1093/mnras/stx385}

\bibitem[{{Jones} {et~al.}(2020){Jones}, {Conroy}, {Horvat}, {Giammarco},
  {Kochoska}, {Pablo}, {Brown}, {Sowicka}, \& {Pr{\v{s}}a}}]{2020Jones}
{Jones}, D., {Conroy}, K.~E., {Horvat}, M., {et~al.} 2020, \apjs, 247, 63,
  \dodoi{10.3847/1538-4365/ab7927}

\bibitem[{{Knigge} {et~al.}(2022){Knigge}, {Toonen}, \&
  {Boekholt}}]{Knigge2022}
{Knigge}, C., {Toonen}, S., \& {Boekholt}, T.~C.~N. 2022, \mnras, 514, 1895,
  \dodoi{10.1093/mnras/stac1336}

\bibitem[{{Kovalev} {et~al.}(2024){Kovalev}, {Chen}, \& {Han}}]{2024Kovalev}
{Kovalev}, M., {Chen}, X., \& {Han}, Z. 2024, \mnras, 527, 346,
  \dodoi{10.1093/mnras/stad3185}

\bibitem[{{Kraus} \& {Hillenbrand}(2009)}]{Hillenbrand2009}
{Kraus}, A.~L., \& {Hillenbrand}, L.~A. 2009, \apj, 704, 531,
  \dodoi{10.1088/0004-637X/704/1/531}

\bibitem[{Li {et~al.}(2021)Li, Shi, Yan, Fu, Li, \& Hou}]{li-2021}
Li, C., Shi, J., Yan, H., {et~al.} 2021, \apjs, 256, 31,
  \dodoi{10.3847/1538-4365/ac22a8}

\bibitem[{Li {et~al.}(2022)Li, Li, Liu, Li, Guo, Wang, Chen, Xing, Hou, \&
  Han}]{Li-2022}
Li, J., Li, J., Liu, C., {et~al.} 2022, The Astrophysical Journal, 933, 119,
  \dodoi{10.3847/1538-4357/ac731d}

\bibitem[{{Li} {et~al.}(2024){Li}, {Liu}, {Luo}, {Zhang}, {Li}, {Li}, {Han},
  {Chen}, {Zhang}, {Wang}, {Fang}, {Xing}, {Zhang}, \& {Jin}}]{2024lijiao}
{Li}, J., {Liu}, C., {Luo}, C., {et~al.} 2024, \apj, 964, 86,
  \dodoi{10.3847/1538-4357/ad20e6}

\bibitem[{{Luo} {et~al.}(2015){Luo}, {Zhao}, {Zhao}, {Deng}, {Liu}, {Jing},
  {Wang}, {Zhang}, {Shi}, {Cui}, {Chu}, {Li}, {Bai}, {Wu}, {Cai}, {Cao}, {Cao},
  {Carlin}, {Chen}, {Chen}, {Chen}, {Chen}, {Chen}, {Chen}, {Chen},
  {Christlieb}, {Chu}, {Cui}, {Dong}, {Du}, {Fan}, {Feng}, {Fu}, {Gao}, {Gong},
  {Gu}, {Guo}, {Han}, {He}, {Hou}, {Hou}, {Hou}, {Hu}, {Hu}, {Hu}, {Huo},
  {Jia}, {Jiang}, {Jiang}, {Jiang}, {Jin}, {Kong}, {Kong}, {Lei}, {Li}, {Li},
  {Li}, {Li}, {Li}, {Li}, {Li}, {Li}, {Li}, {Li}, {Li}, {Li}, {Liang}, {Lin},
  {Liu}, {Liu}, {Liu}, {Liu}, {Lu}, {Luo}, {Mao}, {Newberg}, {Ni}, {Qi}, {Qi},
  {Shen}, {Shi}, {Song}, {Song}, {Su}, {Su}, {Tang}, {Tao}, {Tian}, {Wang},
  {Wang}, {Wang}, {Wang}, {Wang}, {Wang}, {Wang}, {Wang}, {Wang}, {Wang},
  {Wang}, {Wang}, {Wang}, {Wang}, {Wang}, {Wang}, {Wang}, {Wang}, {Wang},
  {Wang}, {Wei}, {Wei}, {Wu}, {Wu}, {Wu}, {Wu}, {Xing}, {Xu}, {Xu}, {Xu},
  {Yan}, {Yang}, {Yang}, {Yang}, {Yang}, {Yao}, {Yu}, {Yuan}, {Yuan}, {Yuan},
  {Yuan}, {Zhai}, {Zhang}, {Zhang}, {Zhang}, {Zhang}, {Zhang}, {Zhang},
  {Zhang}, {Zhang}, {Zhao}, {Zhou}, {Zhou}, {Zhu}, {Zhu}, {Zou}, \&
  {Zuo}}]{2015luo}
{Luo}, A.~L., {Zhao}, Y.-H., {Zhao}, G., {et~al.} 2015, Research in Astronomy
  and Astrophysics, 15, 1095, \dodoi{10.1088/1674-4527/15/8/002}

\bibitem[{{Merkle}(1988)}]{Sahlmann2011}
{Merkle}, F., ed. 1988, European Southern Observatory Conference and Workshop
  Proceedings, Vol.~29, {High-resolution imaging by interferometry}

\bibitem[{{Merle} {et~al.}(2017){Merle}, {Van Eck}, {Jorissen}, {Van der
  Swaelmen}, {Masseron}, {Zwitter}, {Hatzidimitriou}, {Klutsch}, {Pourbaix},
  {Blomme}, {Worley}, {Sacco}, {Lewis}, {Abia}, {Traven}, {Sordo}, {Bragaglia},
  {Smiljanic}, {Pancino}, {Damiani}, {Hourihane}, {Gilmore}, {Randich},
  {Koposov}, {Casey}, {Morbidelli}, {Franciosini}, {Magrini}, {Jofre},
  {Costado}, {Jeffries}, {Bergemann}, {Lanzafame}, {Bayo}, {Carraro},
  {Flaccomio}, {Monaco}, \& {Zaggia}}]{2017mer}
{Merle}, T., {Van Eck}, S., {Jorissen}, A., {et~al.} 2017, \aap, 608, A95,
  \dodoi{10.1051/0004-6361/201730442}

\bibitem[{{Moharana} {et~al.}(2024){Moharana}, {He{\l}miniak}, {Marcadon},
  {Pawar}, {Pawar}, {Konacki}, {Jord{\'a}n}, {Brahm}, \&
  {Espinoza}}]{2024Moharana}
{Moharana}, A., {He{\l}miniak}, K.~G., {Marcadon}, F., {et~al.} 2024, arXiv
  e-prints, arXiv:2405.12136, \dodoi{10.48550/arXiv.2405.12136}

\bibitem[{{Pan} {et~al.}(2022){Pan}, {Fu}, {Zhang}, {Wang}, \& {Li}}]{2022pan}
{Pan}, Y., {Fu}, J.-N., {Zhang}, X.-B., {Wang}, J.-X., \& {Li}, C.-Q. 2022,
  Research in Astronomy and Astrophysics, 22, 075014,
  \dodoi{10.1088/1674-4527/ac712f}

\bibitem[{{Perets} \& {Fabrycky}(2009)}]{Perets2009}
{Perets}, H.~B., \& {Fabrycky}, D.~C. 2009, \apj, 697, 1048,
  \dodoi{10.1088/0004-637X/697/2/1048}

\bibitem[{{Podsiadlowski}(2021)}]{Eisner2022}
{Podsiadlowski}, P. 2021, in American Astronomical Society Meeting Abstracts,
  Vol. 238, American Astronomical Society Meeting Abstracts, 104.02

\bibitem[{{Pr{\v{s}}a} \& {Zwitter}(2005)}]{2005Pr}
{Pr{\v{s}}a}, A., \& {Zwitter}, T. 2005, \apj, 628, 426, \dodoi{10.1086/430591}

\bibitem[{{Pr{\v{s}}a} {et~al.}(2016){Pr{\v{s}}a}, {Conroy}, {Horvat}, {Pablo},
  {Kochoska}, {Bloemen}, {Giammarco}, {Hambleton}, \& {Degroote}}]{2016Pr}
{Pr{\v{s}}a}, A., {Conroy}, K.~E., {Horvat}, M., {et~al.} 2016, \apjs, 227, 29,
  \dodoi{10.3847/1538-4365/227/2/29}

\bibitem[{{Raghavan} {et~al.}(2010){Raghavan}, {McAlister}, {Henry}, {Latham},
  {Marcy}, {Mason}, {Gies}, {White}, \& {ten Brummelaar}}]{2010ragha}
{Raghavan}, D., {McAlister}, H.~A., {Henry}, T.~J., {et~al.} 2010, \apjs, 190,
  1, \dodoi{10.1088/0067-0049/190/1/1}

\bibitem[{{Ricker} {et~al.}(2015){Ricker}, {Winn}, {Vanderspek}, {Latham},
  {Bakos}, {Bean}, {Berta-Thompson}, {Brown}, {Buchhave}, {Butler}, {Butler},
  {Chaplin}, {Charbonneau}, {Christensen-Dalsgaard}, {Clampin}, {Deming},
  {Doty}, {De Lee}, {Dressing}, {Dunham}, {Endl}, {Fressin}, {Ge}, {Henning},
  {Holman}, {Howard}, {Ida}, {Jenkins}, {Jernigan}, {Johnson}, {Kaltenegger},
  {Kawai}, {Kjeldsen}, {Laughlin}, {Levine}, {Lin}, {Lissauer}, {MacQueen},
  {Marcy}, {McCullough}, {Morton}, {Narita}, {Paegert}, {Palle}, {Pepe},
  {Pepper}, {Quirrenbach}, {Rinehart}, {Sasselov}, {Sato}, {Seager},
  {Sozzetti}, {Stassun}, {Sullivan}, {Szentgyorgyi}, {Torres}, {Udry}, \&
  {Villasenor}}]{Ricker2015}
{Ricker}, G.~R., {Winn}, J.~N., {Vanderspek}, R., {et~al.} 2015, Journal of
  Astronomical Telescopes, Instruments, and Systems, 1, 014003,
  \dodoi{10.1117/1.JATIS.1.1.014003}

\bibitem[{{Sana} \& {Evans}(2011)}]{Evans2011}
{Sana}, H., \& {Evans}, C.~J. 2011, in Active OB Stars: Structure, Evolution,
  Mass Loss, and Critical Limits, ed. C.~{Neiner}, G.~{Wade}, G.~{Meynet}, \&
  G.~{Peters}, Vol. 272, 474--485, \dodoi{10.1017/S1743921311011124}

\bibitem[{{Sana} {et~al.}(2012){Sana}, {de Mink}, {de Koter}, {Langer},
  {Evans}, {Gieles}, {Gosset}, {Izzard}, {Le Bouquin}, \&
  {Schneider}}]{Sana2012}
{Sana}, H., {de Mink}, S.~E., {de Koter}, A., {et~al.} 2012, Science, 337, 444,
  \dodoi{10.1126/science.1223344}

\bibitem[{{Silsbee} \& {Tremaine}(2017)}]{Silsbee2017}
{Silsbee}, K., \& {Tremaine}, S. 2017, \apj, 836, 39,
  \dodoi{10.3847/1538-4357/aa5729}

\bibitem[{{Sullivan} {et~al.}(2015){Sullivan}, {Winn}, {Berta-Thompson},
  {Charbonneau}, {Deming}, {Dressing}, {Latham}, {Levine}, {McCullough},
  {Morton}, {Ricker}, {Vanderspek}, \& {Woods}}]{Sullivan2015}
{Sullivan}, P.~W., {Winn}, J.~N., {Berta-Thompson}, Z.~K., {et~al.} 2015, \apj,
  809, 77, \dodoi{10.1088/0004-637X/809/1/77}

\bibitem[{{Tokovinin}(2014)}]{2014AJTokovinin}
{Tokovinin}, A. 2014, \aj, 147, 87, \dodoi{10.1088/0004-6256/147/4/87}

\bibitem[{{Tokovinin}(2018)}]{2018tokovin}
---. 2018, \apjs, 235, 6, \dodoi{10.3847/1538-4365/aaa1a5}

\bibitem[{{Tokovinin} \& {Moe}(2020)}]{2019Tokovinin&Moe}
{Tokovinin}, A., \& {Moe}, M. 2020, \mnras, 491, 5158,
  \dodoi{10.1093/mnras/stz3299}

\bibitem[{{Tokovinin} {et~al.}(2006){Tokovinin}, {Thomas}, {Sterzik}, \&
  {Udry}}]{Tokovinin2006}
{Tokovinin}, A., {Thomas}, S., {Sterzik}, M., \& {Udry}, S. 2006, \aap, 450,
  681, \dodoi{10.1051/0004-6361:20054427}

\bibitem[{{Toonen} {et~al.}(2022){Toonen}, {Boekholt}, \& {Portegies
  Zwart}}]{2022tooten}
{Toonen}, S., {Boekholt}, T.~C.~N., \& {Portegies Zwart}, S. 2022, \aap, 661,
  A61, \dodoi{10.1051/0004-6361/202141991}

\bibitem[{{Toonen} {et~al.}(2016){Toonen}, {Hamers}, \& {Portegies
  Zwart}}]{Toonen2017}
{Toonen}, S., {Hamers}, A., \& {Portegies Zwart}, S. 2016, Computational
  Astrophysics and Cosmology, 3, 6, \dodoi{10.1186/s40668-016-0019-0}

\bibitem[{{VanderPlas}(2017)}]{2017Lomb}
{VanderPlas}, J.~T. 2017, ArXiv e-prints.
\newblock \doarXiv{1703.09824}

\bibitem[{{Wilson} \& {Devinney}(1971)}]{1971Wilson}
{Wilson}, R.~E., \& {Devinney}, E.~J. 1971, \apj, 166, 605,
  \dodoi{10.1086/150986}

\bibitem[{{Xiong} {et~al.}(2024){Xiong}, {Ding}, {Li}, {Ge}, {Cheng}, {Ji},
  {Han}, \& {Chen}}]{2024Xiong}
{Xiong}, J., {Ding}, X., {Li}, J., {et~al.} 2024, \apjs, 270, 20,
  \dodoi{10.3847/1538-4365/ad0ceb}

\bibitem[{{Xiong} {et~al.}(2023){Xiong}, {Liu}, {Li}, {Li}, {Zhang}, {Chen},
  {Luo}, {Cao}, \& {Zhao}}]{2023Xiong}
{Xiong}, J., {Liu}, C., {Li}, J., {et~al.} 2023, \aj, 165, 30,
  \dodoi{10.3847/1538-3881/aca288}

\bibitem[{{Zhang} {et~al.}(2020){Zhang}, {Liu}, \& {Deng}}]{2020zhangbo}
{Zhang}, B., {Liu}, C., \& {Deng}, L.-C. 2020, \apjs, 246, 9,
  \dodoi{10.3847/1538-4365/ab55ef}

\bibitem[{{Zhang} {et~al.}(2021{\natexlab{a}}){Zhang}, {Li}, {Yang}, {Xiong},
  {Fu}, {Liu}, {Tian}, {Li}, {Wang}, {Liang}, {Zhou}, {Zong}, {Yang}, {Liu}, \&
  {Hou}}]{2021zhangbo}
{Zhang}, B., {Li}, J., {Yang}, F., {et~al.} 2021{\natexlab{a}}, \apjs, 256, 14,
  \dodoi{10.3847/1538-4365/ac0834}

\bibitem[{{Zhang} {et~al.}(2021{\natexlab{b}}){Zhang}, {Li}, {Yang}, {Xiong},
  {Fu}, {Liu}, {Tian}, {Li}, {Wang}, {Liang}, {Zhou}, {Zong}, {Yang}, {Liu}, \&
  {Hou}}]{2021ApJS}
---. 2021{\natexlab{b}}, \apjs, 256, 14, \dodoi{10.3847/1538-4365/ac0834}

\bibitem[{{Zhao} {et~al.}(2012){Zhao}, {Zhao}, {Chu}, {Jing}, \&
  {Deng}}]{Zhao2012}
{Zhao}, G., {Zhao}, Y.-H., {Chu}, Y.-Q., {Jing}, Y.-P., \& {Deng}, L.-C. 2012,
  Research in Astronomy and Astrophysics, 12, 723,
  \dodoi{10.1088/1674-4527/12/7/002}

\end{thebibliography}

\appendix

\begin{figure}[ht!]
\centering
\includegraphics[width=11cm]{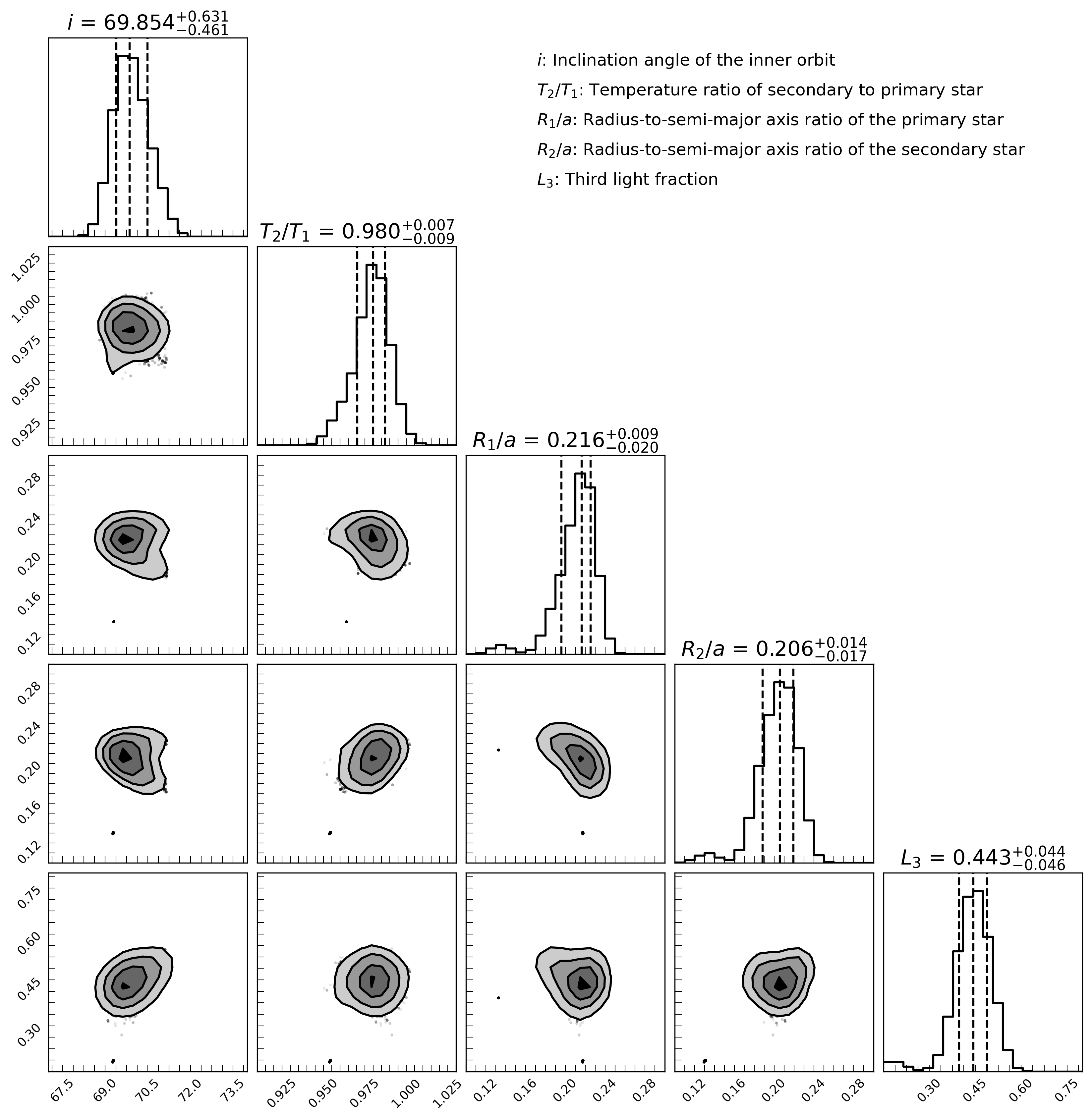}
\caption{The MCMC corner plot for the triple star system $Gaia$ DR3 249662295687401216. The fitted model includes the binary parameters (inclination incl $i$, temperature ratio $T_2/T_1$, $R_1/a$, $R_2/a$) and the third light fraction $L_3$.
\label{figure 6}}
\end{figure}

\begin{figure}[ht!]
\centering
\includegraphics[width=11cm]{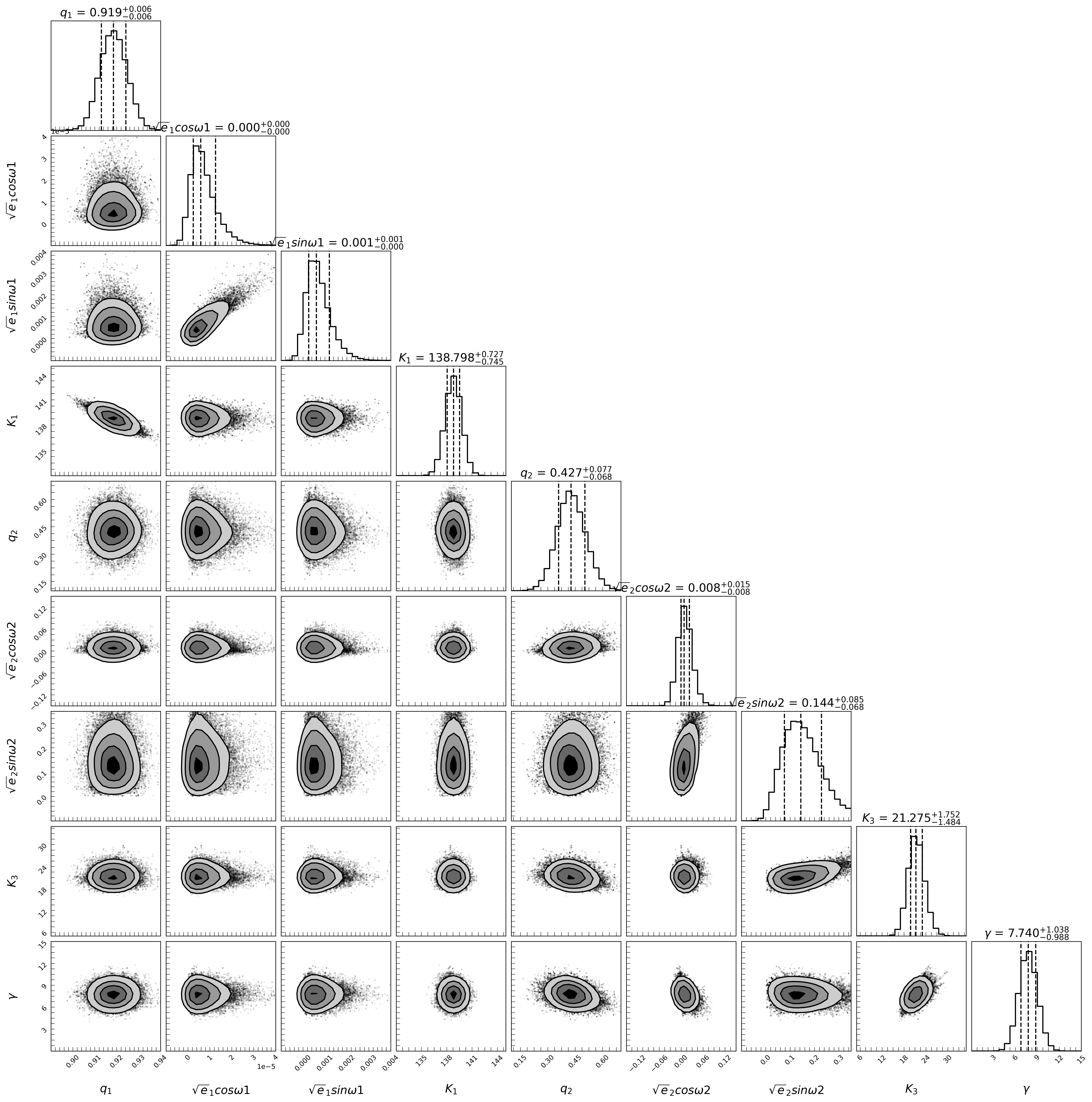}
\caption{MCMC corner plot for the triple-star system $Gaia$ DR3 249662295687401216. The fitted model parameters are: for the nner binary: \(q_1 = 0.919 \pm 0.006\), \(K_1 = 138.798^{+0.727}_{-0.749}\), \(\sqrt{e_1}\cos\omega_1 = 0\), \(\sqrt{e_1}\sin\omega_1 = 0.0014 \pm 0.001\); Outer binary: \(\sqrt{e_2}\cos\omega_2 = -0.008^{+0.015}_{-0.008}\), \(\sqrt{e_2}\sin\omega_2 = 0.144^{+0.085}_{-0.068}\), \(q_2 = 0.0.427^{+0.077}_{-0.068}\), \(K_3 = 21.275^{+1.752}_{-1.484}\); and the system velocity \(\gamma = 7.74 ^{+1.038}_{-0.988} \ \text{km/s}\).
\label{figure m1}}
\end{figure}

\begin{table}[ht]
\centering
\caption{Additional information on the triple $Gaia$ DR3 249662295687401216 in figure 3}
\label{tab4}
\begin{tabular}{ccccccc}
\toprule
BJD & $V_{1}$ & $V_{2}$ & $V_{3}$ & $V_{1err}$ & $V_{2err}$ & $V_{3err}$ \\
\midrule
2459619.96045 & -119.0387 & -7.5837 & 139.5830 & 0.9448 & 1.7338 & 1.6576 \\
2458806.20085 & -105.9431 & -27.5919 & 125.7469 & 1.9054 & 1.1076 & 0.7461 \\
2459158.20122 & -117.1104 & 12.4418 & 142.4093 & 1.6585 & 2.4588 & 2.0134 \\
2459158.16858 & -117.5815 & 12.4548 & 145.9287 & 1.2957 & 3.5850 & 3.8103 \\
2459158.18525 & -117.7914 & 12.8992 & 141.4996 & 0.7086 & 2.1714 & 2.1122 \\
2459565.06937 & -136.8771 & -17.3657 & 131.0411 & 2.4024 & 1.4478 & 1.4987 \\
2459565.09992 & -137.3453 & -17.3151 & 131.1930 & 1.2994 & 1.8312 & 1.8779 \\
\midrule
2459592.96500 & -112.3286 &  & 67.0544 & 2.3578 &  & 2.3612 \\
\midrule
2459150.25300 & 12.0631 &  &  & 1.8233 &  & \\
2459150.23700 & 12.9244 &  &  & 0.6829 &  & \\
2459157.17700 & 9.8588 &  &  & 0.9247 &  & \\
2459162.23300 & 11.4177 &  &  & 1.7500 &  & \\
2459150.27000 & 9.8714 &  &  & 2.6346 &  & \\
2459157.19400 & 10.7534 &  &  & 2.0259 &  & \\
\bottomrule
\end{tabular}
\end{table}

\begin{table}[ht]
\centering
\caption{ Additional information on the triple $Gaia$ DR3 2077667962475652864}
\label{tab5}
\scalebox{0.99}{
\begin{tabular}{ccccccc}
\toprule
\textbf{BJD} & \textbf{$V_1$} & \textbf{$V_{1err}$} & \textbf{$V_2$} & \textbf{$V_{2err}$} & \textbf{$V_3$} & \textbf{$V_{3err}$} \\ 
\toprule
2458625.27818 & -71.5141 & 1.2114 & 6.5813 & 0.7544 & 88.1354 & 1.2937 \\ 
2458625.29415 & -72.8183 & 0.6400 & 6.8530 & 0.9039 & 89.2929 & 1.7148 \\ 
2458625.31082 & -74.7479 & 1.2145 & 8.2974 & 1.1873 & 89.6951 & 2.9073 \\ 
2458644.23239 & -85.2757 & 1.5887 & 11.7181 & 0.6786 & 90.1328 & 1.2854 \\
2458644.24906 & -86.1013 & 2.0255 & 11.1990 & 0.9110 & 90.2436 & 1.0266 \\
2458644.26503 & -86.6967 & 1.8939 & 11.6369 & 0.9420 & 90.3599 & 1.0032 \\
2458644.28170 & -87.3995 & 1.8057 & 10.8190 & 0.8663 & 90.1659 & 1.5193 \\
2458646.25191 & -74.0192 & 1.6597 & 11.8710 & 0.6789 & 87.0328 & 1.6974 \\
2458646.26857 & -73.4940 & 1.2577 & 12.1782 & 0.8371 & 85.7476 & 2.2166 \\
2458646.29080 & -72.5440 & 1.0093 & 11.3691 & 1.3673 & 83.5492 & 2.1834 \\
2459003.22660 & -69.2077 & 0.8406 & -3.0762 & 1.9225 & 86.4226 & 1.7721 \\
2459003.24258 & -70.1571 & 1.4678 & -3.6965 & 2.1558 & 87.1333 & 1.6762 \\
2459003.25855 & -71.6484 & 0.8941 & -4.5256 & 1.8252 & 88.0940 & 1.7160 \\
2459003.27522 & -73.1173 & 1.0875 & -4.7221 & 1.9872 & 89.3608 & 1.6604 \\
2459003.29119 & -74.6977 & 0.9365 & -3.8181 & 2.2706 & 89.9883 & 1.8310 \\
2459003.30716 & -76.0505 & 1.2940 & -4.3444 & 1.7029 & 90.9279 & 1.5418 \\
2459015.20619 & -66.7174 & 1.2193 & -0.9132 & 4.4873 & 89.7898 & 2.0159 \\
2459015.22216 & -68.1226 & 0.9288 & -1.4899 & 3.8154 & 90.6656 & 2.0084 \\
2459015.23883 & -69.2202 & 1.0621 & -0.6591 & 3.4082 & 91.5898 & 1.8986 \\
2459015.26175 & -71.3216 & 1.0463 & -1.3636 & 2.5738 & 93.0927 & 1.7058 \\
2459015.27842 & -72.3435 & 0.8253 & -1.6658 & 3.2843 & 94.0681 & 1.7327 \\
2459015.29439 & -73.6048 & 1.1894 & -1.6872 & 2.9163 & 94.6506 & 1.6963 \\
\midrule
2459004.26067 & 7.7075 & 1.0468 & - & - & - & - \\ 
2459004.30928 & 6.7273 & 1.2495 & - & - & - & - \\ 
2459016.29859 & 6.0380 & 1.2742 & - & - & - & - \\ 
2459004.27665 & 7.4306 & 1.1597 & - & - & - & - \\ 
2459011.31162 & 5.3659 & 0.9330 & - & - & - & - \\ 
2458267.31529 & 0.3463 & 1.0208 & - & - & - & - \\ 
2459004.29331 & 7.2051 & 1.2166 & - & - & - & - \\ 
2458267.26946 & -0.6726 & 1.1026 & - & - & - & - \\
2459011.26231 & 7.2487 & 1.1327 & - & - & - & - \\ 
2459016.26595 & 6.6073 & 1.2248 & - & - & - & - \\ 
2459004.24470 & 7.9278 & 0.9776 & - & - & - & - \\ 
2458267.23126 & -1.0669 & 1.1936 & - & - & - & - \\
2459011.29495 & 5.8890 & 0.9332 & - & - & - & - \\ 
2458267.25696 & -0.7937 & 1.1488 & - & - & - & - \\
2459016.28192 & 6.3168 & 1.2361 & - & - & - & - \\ 
2459016.24997 & 6.4894 & 1.1962 & - & - & - & - \\ 
2458267.28196 & -0.1141 & 1.0828 & - & - & - & - \\
2458267.29515 & -0.0591 & 0.9924 & - & - & - & - \\
2459011.27898 & 6.4819 & 1.0363 & - & - & - & - \\ 
2458267.24376 & -1.0640 & 1.1436 & - & - & - & - \\
\bottomrule
\end{tabular}}
\end{table}

\begin{figure}[ht!]
\centering
\includegraphics[width=11cm]{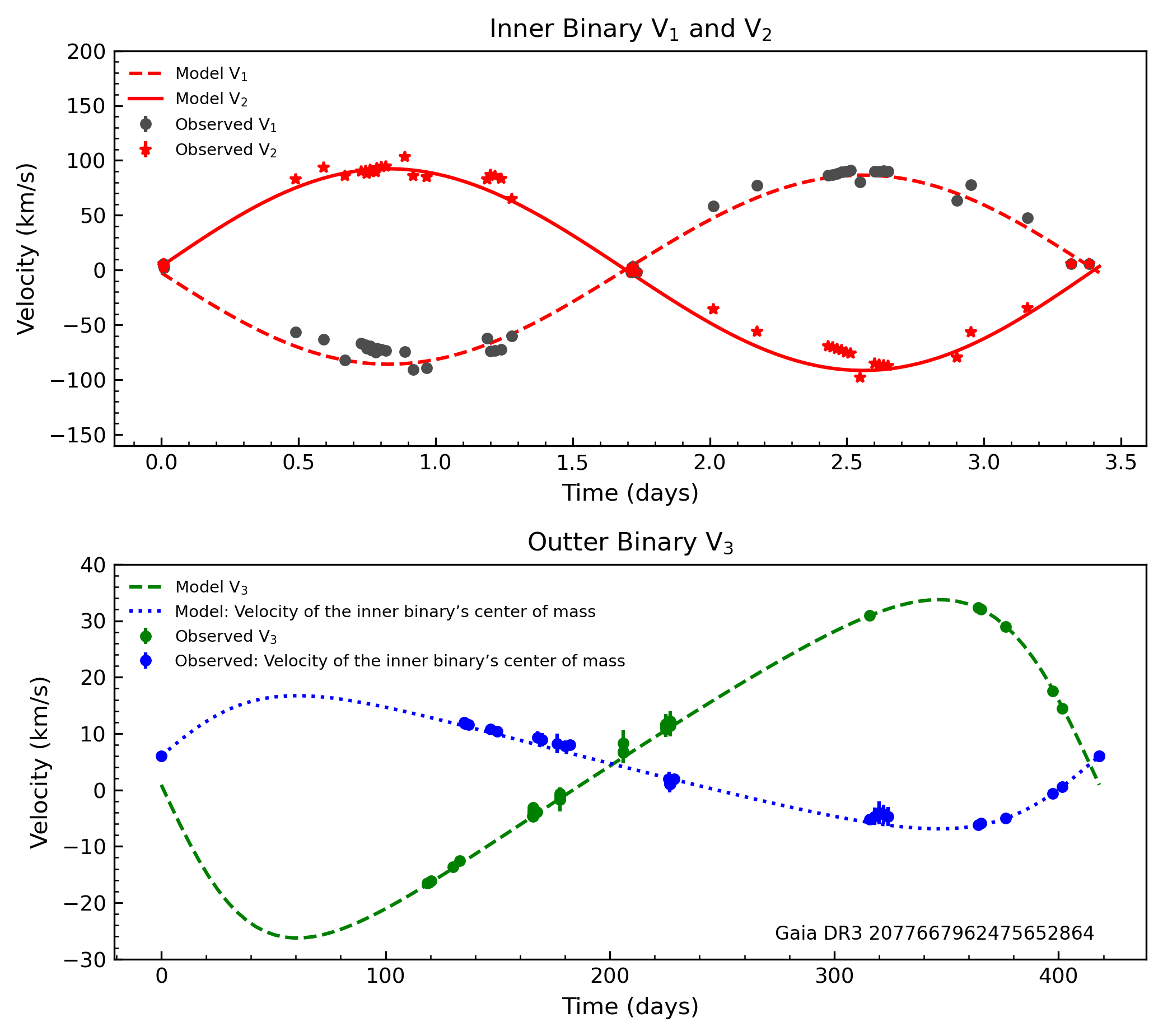}
\caption{RV fitting for the triple star system $Gaia$ DR3 2077667962475652864.The black points and red stars represent the observed data after phase folding, and the red line indicates the fitted RV curve. The blue points show the velocity of the inner binary’s center of mass, while the green dashed line represents the fitted curve for the inner binary’s center of mass velocity. The green points indicate the velocity of the tertiary star, and the green dashed line illustrates the fitted curve for the tertiary star’s velocity. $V_{1}$ and $V_{2}$ denote the RVs of the primary and secondary stars in the inner binary, respectively, and $V_{3}$ represents the RV curve of the tertiary star.
\label{figure t1}}
\end{figure} 

\begin{figure}[ht!]
\centering
\includegraphics[width=11cm]{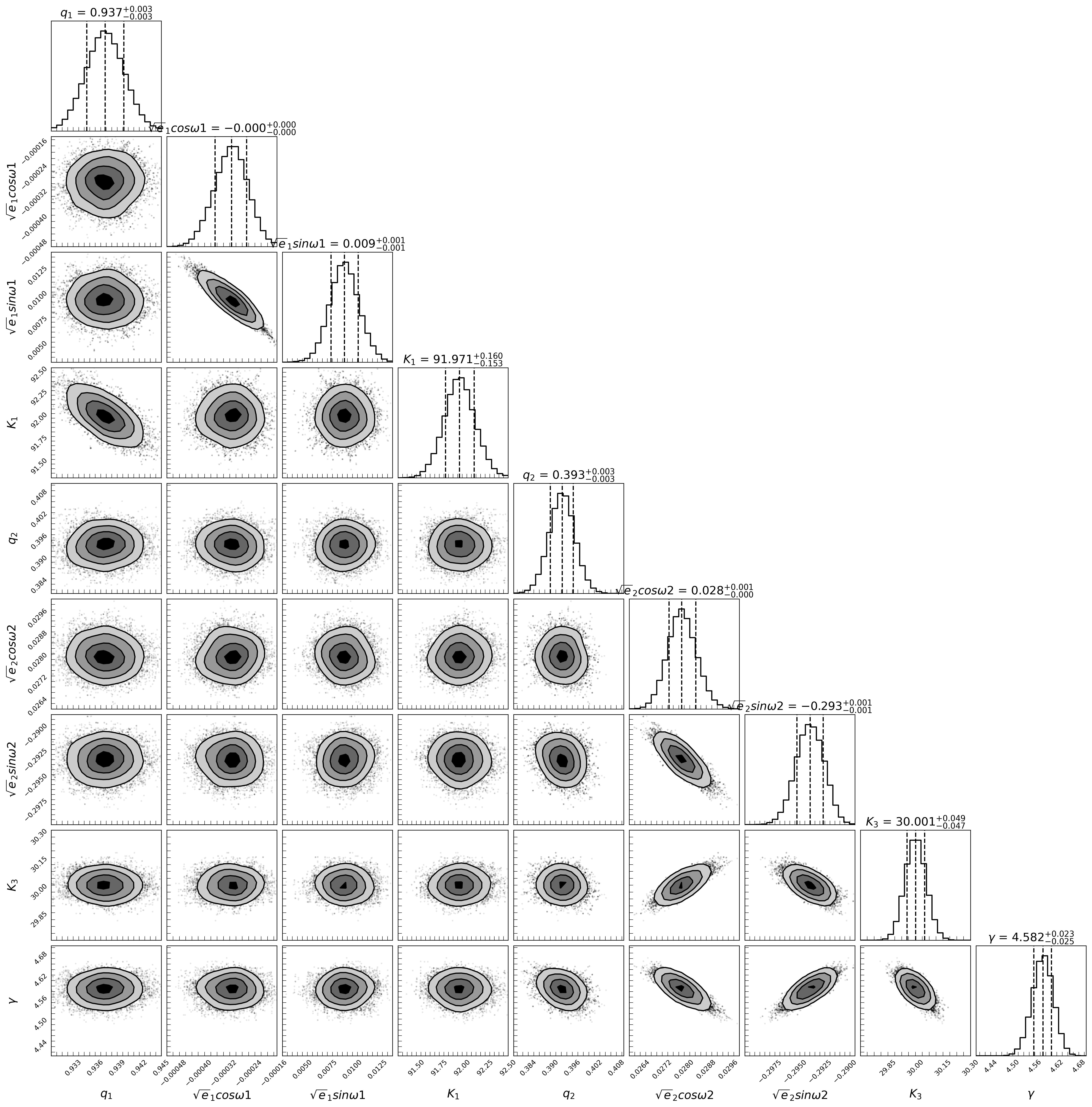}
\caption{MCMC corner plot for the triple-star system $Gaia$ DR3 2077667962475652864. The fitted model parameters are: for the inner binary, \( q_1 = 0.937 \pm 0.003  \), \( K_1 = 91.971^{+0.160}_{-0.153} \), \( \sqrt{e}_1 \cos \omega_1 = 0 \), \( \sqrt{e}_1 \sin \omega_1 = 0.009 \pm 0.001 \); for the outer binary, \( \sqrt{e}_2 \cos \omega_2 = 0.028^{+0.001}_{-0.0} \), \( \sqrt{e}_2 \sin \omega_2 = -0.293 \pm 0.001 \), \( q_2 = 0.393 \pm 0.003 \), \( K_3 = 30.001^{+0.049}_{-0.047} \); and the systemic velocity \( \gamma = 4.582^{+0.023}_{-0.025} \) km/s.
\label{figure t2}}
\end{figure} 


\begin{figure}[ht!]
\centering
\includegraphics[width=11cm]{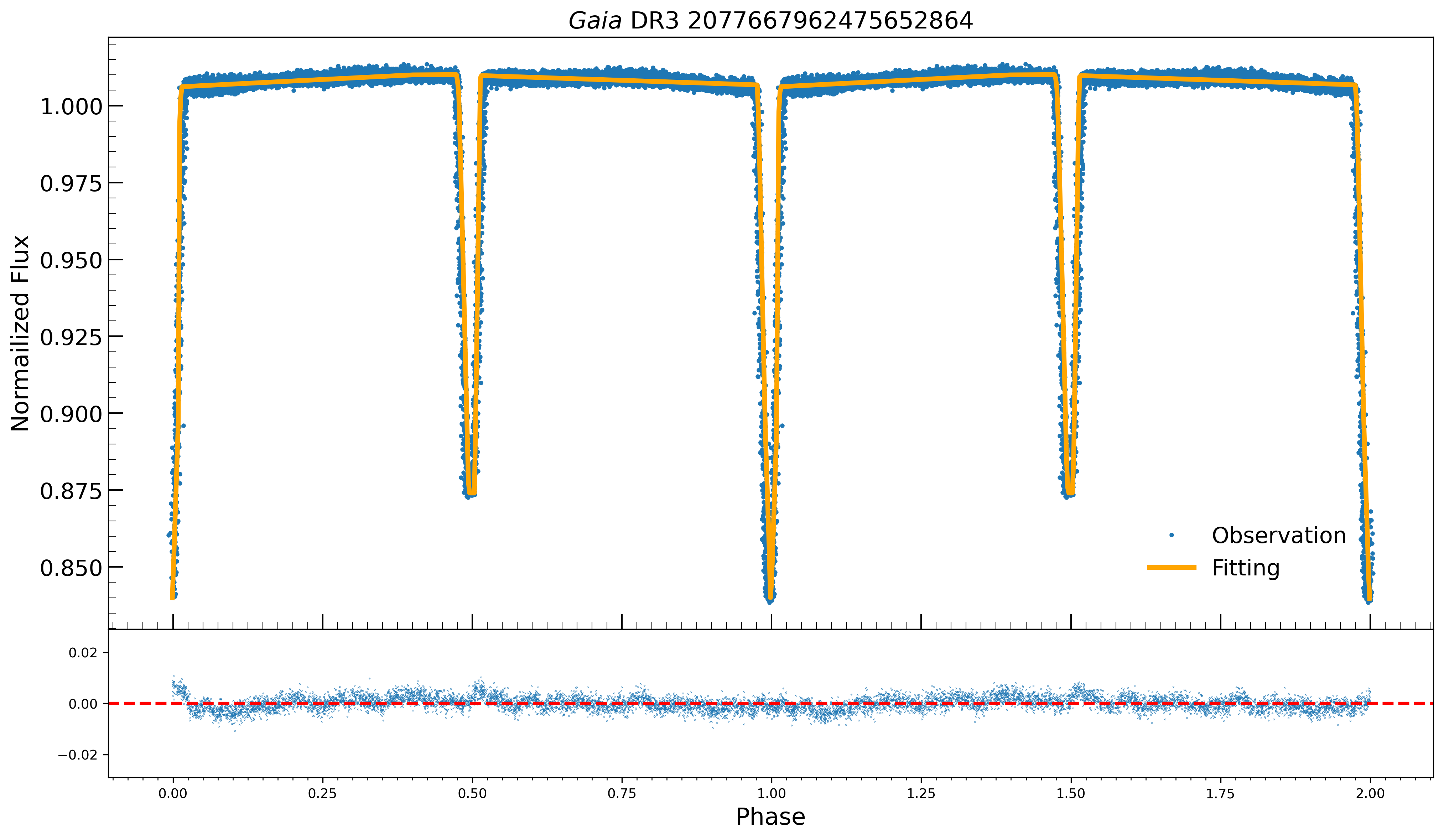}
\caption{ Light curve of the triple $Gaia$ DR3 2077667962475652864. The blue dots indicate phase-folded observations, and the solid yellow line indicates the results of the fit. The horizontal axis indicates the phase, and the fitting residuals are shown at the bottom of the plot.
\label{figure t3}}
\end{figure} 

\begin{figure}[ht!]
\centering
\includegraphics[width=11cm]{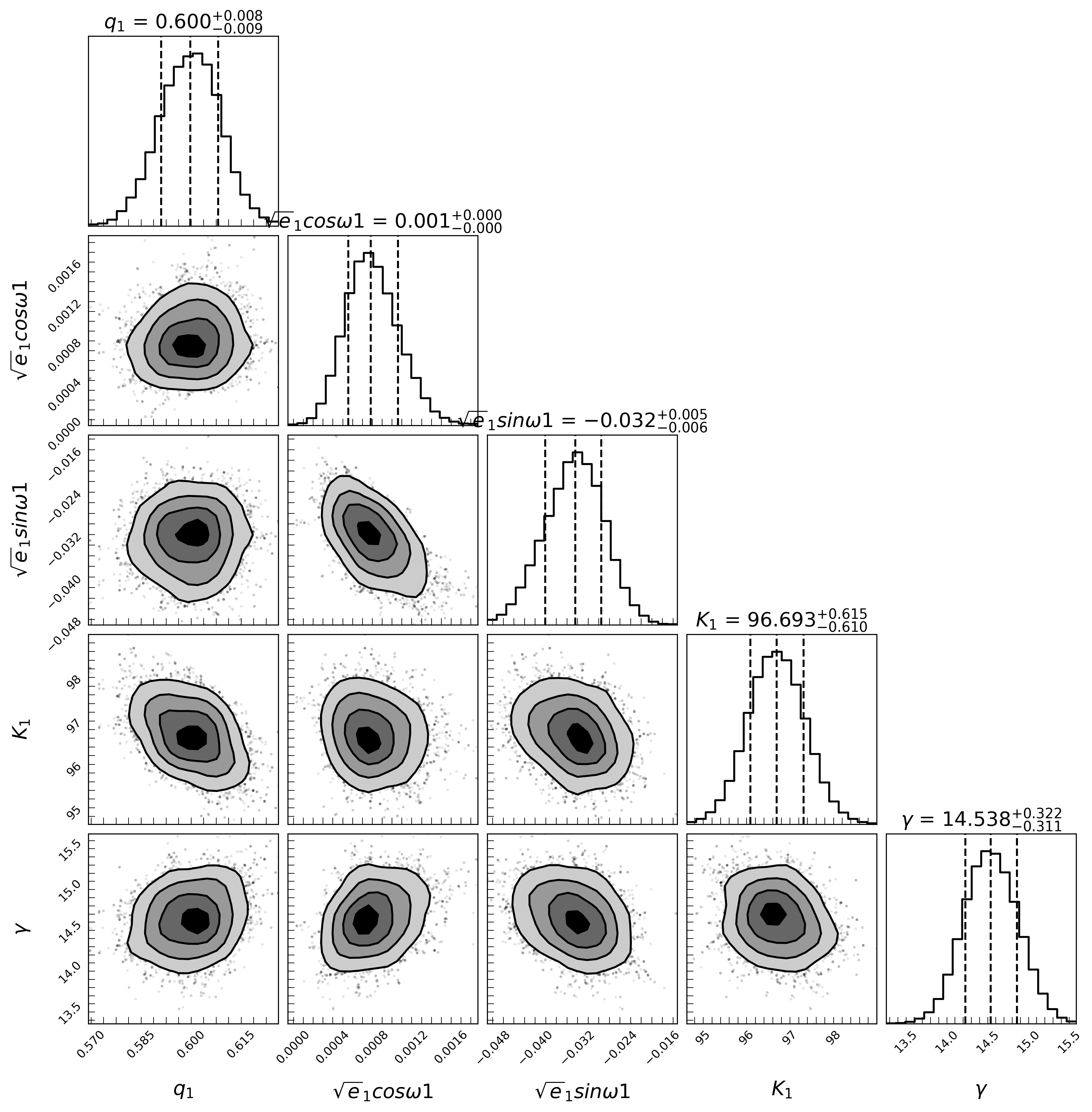}
\caption{The MCMC corner plot for the triple system j03, representing the results of inner orbital fitting. The fitted parameters are: \(q_1 = 0.600 ^{+0.008}_{-0.009}\), \(K_1 = 96.693^{+0.615}_{-0.610}\), \(\sqrt{e_1}\cos\omega_1 = 0.001\), \(\sqrt{e_1}\sin\omega_1 = -0.032^{+0.005}_{-0.006}\) and the  systemic velocity \( \gamma = 14.538^{+0.322}_{-0.311} \) km/s.
\label{figure tt2}}
\end{figure} 

\end{document}